\pdfoutput=1

\documentclass[11pt,twoside,a4paper,cmspaper,final,collab]{cms-tdr}
\def\svnVersion{237267}\def\cmsCernNo{2013-196}\def\cmsCernDate{\today}\def\cmsMessage{Published in Physical Review C as \href{http://dx.doi.org/10.1103/PhysRevC89.044906}{\doi{10.1103/PhysRevC89.044906.}}}

\begin{document}\cmsNoteHeader{HIN-11-005}

\hyphenation{had-ron-i-za-tion}
\hyphenation{cal-or-i-me-ter}
\hyphenation{de-vices}

\RCS$Revision: 164222 $
\RCS$HeadURL: svn+ssh://ssanders@svn.cern.ch/reps/tdr2/papers/HIN-11-005/trunk/HIN-11-005.tex $
\RCS$Id: HIN-11-005.tex 164222 2013-01-10 22:46:58Z ssanders $
\providecommand{\re}{\ensuremath{\cmsSymbolFace{e}}}
\providecommand{\sNN}{\ensuremath{\sqrt{\smash[b]{s_{_\mathrm{NN}}}}}\xspace}
\newlength\cmsFigWidth
\ifthenelse{\boolean{cms@external}}{\setlength\cmsFigWidth{0.48\textwidth}}{\setlength\cmsFigWidth{0.75\textwidth}}
\newlength\cmsFigWidthStd\setlength\cmsFigWidthStd{0.48\textwidth}
\newlength\cmsFigWidthStdWide\ifthenelse{\boolean{cms@external}}{\setlength\cmsFigWidthStdWide{0.85\textwidth}}{\setlength\cmsFigWidthStdWide{0.75\textwidth}}
\cmsNoteHeader{HIN-11-005} 
\title{Measurement of higher-order harmonic azimuthal anisotropy in PbPb collisions at \texorpdfstring{$\sNN = 2.76\TeV$}{sqrt[sNN]=2.76 TeV}}

\date{\today}

\abstract{Measurements are presented by the CMS Collaboration at the Large Hadron Collider (LHC) of the higher-order
harmonic coefficients that describe the azimuthal anisotropy of charged particles emitted in
$\sNN = 2.76$\TeV PbPb collisions.
Expressed in terms of the Fourier components of the azimuthal
distribution, the n = 3--6  harmonic coefficients are presented for charged particles as a function of
their transverse momentum ($0.3<\pt<8.0\GeVc$), collision centrality (0--70\%), and
pseudorapidity ($\abs{\eta} < 2.0$). The data are analyzed using the event plane, multi-particle cumulant, and Lee--Yang zeros methods, which provide different sensitivities to initial state fluctuations.   Taken together with earlier LHC measurements of elliptic flow ($n=2$), the results on higher-order harmonic coefficients  develop a more complete picture of the collective motion in
high-energy heavy-ion collisions and shed light on the properties of the
produced medium.}

\hypersetup{%
pdfauthor={CMS Collaboration},%
pdftitle={Measurement of higher-order harmonic azimuthal anisotropy in PbPb collisions at a nucleon-nucleon
center-of-mass energy of 2.76 TeV},%
pdfsubject={CMS},%
pdfkeywords={CMS, physics, heavy ions, higher harmonic flow, eccentricity scaling}}

\maketitle

\section{Introduction}
\label{sec:Introduction}
In the collision of two heavy ions moving relativistically, a high-density energetic state of matter
is created in the overlap region of the two Lorentz-contracted nuclei.
Earlier studies at the Relativistic Heavy-Ion Collider (RHIC),
where gold nuclei were collided at nucleon-nucleon center-of-mass energies up to
$\sNN=200\GeV$~\cite{BRAHMS,PHOBOS,STAR,PHENIX}, found that the particles produced in rare, high-momentum-transfer scatterings encounter a dense medium with high stopping power for colored probes.  The low-momentum particles that comprise the bulk of the medium exhibit strong azimuthal anisotropies that indicate a collective fluid expansion.  These findings have been interpreted
as manifestations of a strongly interacting quark-gluon plasma.
The created medium is found to
behave as a nearly perfect fluid~\cite{PhysRevC.78.034915,PhysRevLett.97.162302,PhysRevLett.98.172301,
PhysRevC.76.024905} with a ratio of shear viscosity $\eta$ to entropy density
$s$ approaching the conjectured lower limit of ${\eta }/{s} \ge {\hbar }/{{(4\pi {k_B})}}$, found by considering the implications of the Heisenberg uncertainty principle for a
viscous plasma~\cite{PhysRevLett.94.111601}. Pressure gradients that develop in the fluid during the
collision result in an anisotropic momentum distribution of the outflowing matter, which, in
turn, leads to a preferential emission of particles in the short direction
of the lenticular-shaped overlap
region~\cite{PhysRevD.46.229,PhysRevLett.82.2048,Reisdorf:1997fx}.
The hydrodynamic behavior suggests that local thermal equilibrium may be
achieved very rapidly in the hot medium,
with the observed anisotropy in particle emission therefore being
sensitive to the basic properties of the created system, such as
the equation of state, the $\eta/s$ value, and the speed of sound in the
medium.  The anisotropy also depends on the initial conditions, allowing the investigation of whether a Glauber-like picture of
individual nucleon collisions~\cite{AnnRevNuclPart.57.205} prevails  or if gluon
saturation effects, as found in the color glass condensate (CGC) model~\cite{McLerran:2001sr, PhysRevC.86.034908}, play an important  role.

More recently, at the Large Hadron Collider, the azimuthal anisotropy measurements have been
extended to a much higher energy,
with PbPb collisions at  $\sNN=2.76\TeV$~\cite{PhysRevLett.105.252302,
PhysLettB.707.330,PhysRevC.87.014902}.
Moreover, the azimuthal distributions are
being investigated with greater precision with the exploration of
higher-order anisotropies at both the RHIC~\cite{PhysRevC.72.014904,
PhysRecC.75.054906,PhysRevLett.92.062301,PhysRevLett.105.062301,PhysRevLett.107.252301} and
LHC~\cite{PhysLettB.708.249,PhysRevLett.107.032301,PhysRevC.86.014907} facilities.
The azimuthal dependence of the particle yield $N$ can be written in terms of
harmonic expansion coefficients $v_n$; with \cite{PhysRevC.58.1671}

\ifthenelse{\boolean{cms@external}}
{
\begin{multline}
\label{eqn:Eqn1}
E\frac{{\rd^3 N}}{{\rd{}p^3}} = \\
\frac{1}{{2\pi }}\frac{{\rd^2 N}}{{\pt\, \rd\pt\, \rd{}y}}\left( {1 + \sum\limits_{n = 1}^\infty  {2v_n \cos \left[ {n\left( {\phi  - \Psi } \right)} \right]} } \right),
\end{multline}
}
{
\begin{equation}
\label{eqn:Eqn1}
E\frac{{\rd^3 N}}{{\rd{}p^3}} =
\frac{1}{{2\pi }}\frac{{\rd^2 N}}{{\pt\, \rd\pt\, \rd{}y}}\left( {1 + \sum\limits_{n = 1}^\infty  {2v_n \cos \left[ {n\left( {\phi  - \Psi } \right)} \right]} } \right),
\end{equation}
}
where  $\phi$, $E$, $y$, and $\pt$  are the azimuthal angle, energy, rapidity,
and transverse momentum of the particles, respectively.
In analyzing an experimental distribution, the reference angle $\Psi$ needs to be determined empirically and typically corresponds to
the direction of the greatest azimuthal density of outgoing particles.
The higher-order harmonics are expected to be particularly sensitive to fluctuations in the initial conditions~\cite{PhysRevC.81.034915,PhysRevC.77.014906,PhysRevC.81.054905,PhysLettB.641.260,PhysRevC.76.054905,PhysRevC.85.014905,PhysRevC.81.014901,PhysRevC.84.064907,PhysRevC.80.014904,PhysRevC.82.064903,PhysRevC.84.024911,PhysLettB.659.537,PhysRevC.82.034913}
and to the shear viscosity of the created medium~\cite{PhysRevLett.106.042301,PhysRevC.82.034913,PhysRevC.84.044912,PhysRevC.85.024901,PhysRevC.85.034902}.

This paper presents results from the Compact Muon Solenoid (CMS) Collaboration on higher-order harmonic anisotropy components
for PbPb collisions at $\sNN=2.76\TeV$ using the event plane~\cite{PhysRevC.58.1671}, multiparticle cumulant~\cite{PhysRevC.64.054901},
and  Lee--Yang zeros~\cite{NuclPhysA.727.373, Borghini:2004ke} methods to exploit the different sensitivities of these methods to initial state
fluctuations. In
addition to correlations resulting from flow and initial state fluctuations, there exist other sources of azimuthal
correlations, such as those arising from resonance decays and jets.  These ``nonflow" correlations
either do not or only weakly depend on the bulk motion of the medium and need to be
considered when determining the ``true," global collective flow behavior.  The methods discussed in this paper
are affected differently by the nonflow effects and steps are taken to minimize the influence of these processes when possible.

This work extends the previously published CMS results on elliptic flow
(the $n=2$ harmonic)~\cite{PhysRevC.87.014902}.  The data and event selection used here are
identical to those employed in the elliptic-flow analysis, and the current discussion of the
experimental methods summarizes a more extensive discussion found in the earlier
paper.  New results are presented for harmonics $n=3$, 4, 5, and 6 of charged particle distributions as a function of their
transverse momentum ($0.3<\pt<8.0\GeVc$), collision centrality (0--70\%),  and pseudorapidity ($\abs{\eta}<2.0$). The pseudorapidity is defined in terms of the polar angle $\theta$ with $\eta=-\ln[\tan(\theta/2)]$. The collision centrality reflects the degree of overlap of the two colliding nuclei, with 0\% central events corresponding to impact parameter (i.e., the distance between the centers of the two colliding nuclei at closest approach) $b$ = 0.  Some of the earlier CMS elliptic-flow results are included in order to present a more complete description of the anisotropy behavior. The CMS Collaboration results on higher-order harmonic anisotropies obtained using the two particle correlation technique~\cite{Eur.Phys.J.C72.2012} are also included here for comparison.

The paper is organized as follows.  Section~\ref{sec:ExperimentalMethod} presents an overview of the experimental procedures and different methods
that are used in the analysis.  Various sources of systematic uncertainties are discussed.
This section also develops the Glauber model eccentricities used in
discussing the experimental results.  Section~\ref{sec:results} presents
the ``differential" and yield-weighted average harmonic coefficients (``integral" flow)
for the different methods.  The pseudorapidity dependence is presented
for azimuthal anisotropy coefficients evaluated by the event plane method.  Section~\ref{sec:results} also contains a comparison of
the new CMS results
to previously published results of the ALICE and ATLAS Collaborations, as well as lower-energy results obtained by the PHENIX Collaboration at RHIC.
Section~\ref{sec:discussion} presents a discussion of the results and Section~\ref{sec:conclusions} gives an overall summary.

\section{Experimental Details}
\label{sec:ExperimentalMethod}
The measurements were done with the CMS detector
using data obtained in $\sNN=2.76\GeV$ PbPb collisions during the
2010 heavy-ion run at the LHC.   The analysis
uses the same data and techniques as for the elliptic-flow study of
Ref.~\cite{PhysRevC.87.014902}, allowing for a direct comparison with the
results of that study.

\subsection{Experimental setup}
The CMS detector consists of a silicon tracker,
a crystal electromagnetic calorimeter, and a brass/scintillator hadronic
calorimeter housed within a superconducting solenoid 6\unit{m} in diameter
that provides a 3.8\unit{T} magnetic field. Muons are identified
in gas-ionization chambers that are embedded in a steel flux return yoke. The muon information is
not used in the current analysis.
The CMS detector includes extensive charged particle tracking and forward calorimetry.
The inner tracker, consisting of silicon pixel and strip detector modules,
reconstructs the trajectories of charged particles within the pseudorapidity range
$\abs{\eta} < 2.4$. In the forward region, two steel/quartz-fiber
Cherenkov hadron forward (HF) calorimeters cover a pseudorapidity range of
$2.9 < \abs{\eta} < 5.2$. These calorimeters are azimuthally subdivided into $20^{\circ}$
modular wedges and further segmented to form $0.175\times0.175$ $(\Delta\eta \times \Delta\phi)$ ``towers," where the angle $\phi$ is in radians.
A set of scintillator tiles, the beam scintillator counters (BSC), are mounted on the inner side of the HF calorimeters and are used for triggering and beam-halo rejection. The BSCs cover the range $3.23< \abs{\eta} < 4.65$. CMS uses a right-handed coordinate system  where the $x$, $y$, and $z$~axes are aligned with the radius of the LHC ring, the vertical direction, and the counterclockwise-beam direction, respectively, with the origin located at the center of the nominal interaction region.
A more detailed
description of the CMS detector can be found elsewhere~\cite{JINST}. In this
analysis, the azimuthal anisotropy correlations are determined based on the charged particles reconstructed from the tracks in the
silicon tracker.
The event plane analysis also uses information from the HF calorimeters to
establish event planes that have a large separation in pseudorapidity from the
tracks used to determine the anisotropy harmonics.

Track reconstruction is accomplished by starting with a set of
three reconstructed clusters in the inner layers of the silicon pixel detector that  are compatible with a helical trajectory, have a \pt above a minimum value,
and are within a selected region around
the reconstructed primary collision vertex, as determined through an iterative process.
In determining tracks, trajectories with a minimum \pt value of 0.9\GeVc
are first propagated outward through sequential
silicon strip layers using a combinatorial Kalman filter algorithm~\cite{CMSTracking}.
Then, trajectories in the range of $0.2<\pt<1.8$\GeVc
are determined using only signals in the pixel detector, without requiring
continuation of the trajectories to the  silicon strip detector layers.
The reconstructed tracks from both procedures are then merged, removing duplicate tracks by giving preference
to the tracks extending into the silicon strip layers. A track will be ``misreconstructed" if it is generated by the passage
of more than one charged particle or if other spurious signals in the pixel or strip detectors contribute to its determination.
The effect of misreconstructructed tracks is considered in determining the systematic uncertainties for the different methods, as discussed below.

\subsection{Event and track selection}
Minimum bias PbPb events are triggered
by coincident signals from both ends of the CMS detector in either the BSC or HF detectors.
This trigger is required to be in coincidence with the presence of both colliding
ion bunches in the interaction region. Additional offline event selections are applied
in order to obtain a pure sample of inelastic hadronic collisions, thus removing
contamination from noncollision beam backgrounds and from ultra-peripheral
collisions  (UPC) where an electromagnetic interaction leads to the breakup of one or both Pb
nuclei~\cite{PhysRevC.83.041901}. These offline selections include the requirement
of proper timing of the BSC signals from both sides of the detector, a coincidence of at least three
HF towers on each side of the interaction point, with at least 3\GeV energy deposited in each tower,
a reconstructed vertex compatible
with the expected collision region, and the shape of reconstructed clusters from the
pixel detector being consistent with having been produced by particles originating from the
primary collision vertex. These selections are described in more detail in
Ref.~\cite{PhysRevC.84.024906}.

Events used in this analysis are required to have a longitudinal vertex position within
10~cm of the nominal interaction point of the detector in order to consistently measure
charged particle distributions over the tracker rapidity range.
After all selections, 22.6 million
events remain in the final sample, corresponding to an integrated
luminosity of approximately 3\mubinv. This final sample is the same data set used
in the elliptic-flow study and is described in further detail in Ref.~\cite{PhysRevC.87.014902}.

\subsection{Centrality}

The centrality of a collision is a measure of the degree of overlap of the
colliding  ions. Several quantities depend on the centrality and can be used
for its determination. In this analysis, the centrality is based on the total energy deposited in both HF
calorimeters, with the distribution of the
total energy for all events divided into 40 centrality bins,
each representing 2.5\% of the total PbPb interaction cross section. More central events will
have a larger particle multiplicity and therefore greater total energy in the HF calorimeters.
Events falling into adjacent centrality
bins are then combined to form the 5\% and 10\% bins used to present the final
results. The measured charged particle multiplicity distribution does not represent the full interaction cross section because of inefficiencies in the minimum bias trigger
and the event selection. Monte Carlo simulations (MC) are used to estimate the multiplicity
distribution of charged particles in the regions where events are lost because of low trigger efficiency to correct for the inefficiencies. Comparing the simulated
to the measured distribution, we determine the combined efficiency for the minimum bias
trigger and the event selection to be ($97 \pm 3$)\%. Further discussion of the CMS centrality
determination can be found in Ref.~\cite{JHEP08_2011_141}.

\subsection{Analysis methods}
To extract the
$v_n$ coefficients, the event plane~\cite{PhysRevC.58.1671}, the
generating-function-based multiparticle cumulant~\cite{PhysRevC.64.054901},
and the Lee--Yang zeros~\cite{NuclPhysA.727.373,Borghini:2004ke} methods are used,
as described in the next three subsections.  An earlier CMS paper~\cite{Eur.Phys.J.C72.2012} studied
the higher harmonic coefficients using the two particle correlation method.  For completeness, these earlier results are also presented.  The two particle correlation method is similar to a two particle cumulant analysis although, in this case, with the requirement of a gap in pseudorapdity between the particles in a pair.

\subsubsection{Event Plane method}
The event plane method~\cite{PhysRevC.58.1671} measures the azimuthal anisotropy with respect to
an event plane angle of a given  order $m$ that is  determined in a different
pseudorapidity region from that where the anisotropy coefficient $v_n$ is being
measured.
Expressed in terms of the event plane angle $\Psi_m$ corresponding to harmonic $m$,
the azimuthal distribution can be written as~\cite{PhysRevC.58.1671}
\ifthenelse{\boolean{cms@external}}
{
\begin{multline}
\label{eqn:Eqn3}
\frac{{\rd{}N}}{{\rd\left( {\phi  - {\Psi _m}} \right)}} \propto 1 + \sum\limits_{k = 1}^\infty  {2v_{km}^{{ }}} \left\{ {{\Psi _m}} \right\}\cos \left[ {km\left( {\phi  - {\Psi _m}} \right)} \right],
\end{multline}
}
{
\begin{equation}
\label{eqn:Eqn3}
\frac{{\rd{}N}}{{\rd\left( {\phi  - {\Psi _m}} \right)}} \propto 1 + \sum\limits_{k = 1}^\infty  {2v_{km}^{{\text{obs}}}} \left\{ {{\Psi _m}} \right\}\cos \left[ {km\left( {\phi  - {\Psi _m}} \right)} \right],
\end{equation}
}
where $dN$ is the number of charged particles emitted into a differential azimuthal angular range $\rd(\phi-\Psi_m)$, $n=km$, and the dependence of $v_{n}^\text{obs}$ on the event plane harmonic $m$ is explicitly noted. The event plane harmonic $m$ can take any integer value greater than 0.
Generally, when higher-order event planes are considered ($m>2$), only the $k=1$, $n=m$ term is found to be needed to describe the corresponding azimuthal distribution. For $m>1$, it is not possible to describe the correlations corresponding to $n<m$. It has recently been noted~\cite{PhysRevC.86.044908} that participant fluctuations can lead to terms where
$n$ is not an integer multiple of $m$. These mixed harmonic terms will not be studied in this paper.

The method assumes that the event plane angle is a
pseudorapidity independent global observable. For this analysis, event plane
angles are calculated using the measurements of transverse energy obtained with elements of
the azimuthally symmetric HF calorimeters, using
\begin{equation}
\label{eqn:Eqn4}
{\Psi _m} = \frac{1}{m}{\tan ^{ - 1}}\left\{ {\frac{{\left\langle {{w_i}\sin \left( {m{\phi _i}} \right)} \right\rangle }}{{\left\langle {{w_i}\cos \left( {m{\phi _i}} \right)} \right\rangle }}} \right\},
\end{equation}
where $\phi_i$ is the azimuthal position of the $i$th-tower element. The bracket $\langle\rangle$ indicates a sum over tower elements. The transverse energy in each tower {$i$}
is used as the weight $w_i$.
For each fundamental harmonic $m$ we define two event planes
$\Psi_{m}(\mathrm{HF}-)$ ($-5<\eta<-3$) and $\Psi_{m}(\mathrm{HF}+)$ ($3<\eta<5$), corresponding to the
HF calorimeters on either side of the CMS detector. A standard event plane
flattening procedure is employed to avoid having an azimuthal bias
introduced by detector effects~\cite{PhysRevC.58.1671,PhysRevC.87.014902}. The ``flattened'' event plane distributions show no significant azimuthal anisotropy in any harmonic order $m$ when plotted with respect to the laboratory frame.

The differential-anisotropy parameters  $v_{n}(\pt, \eta)$ are then determined with
\begin{equation}
\label{eqn:Eqn5}
{v_{n}^\text{obs}}\left( {{\pt},\eta  < 0} \right)\ =\  \langle\langle\cos \left[ {n\left\{ { {\phi  - {\Psi _m}\left( {\mathrm{HF}+} \right)} } \right\}} \right]\rangle\rangle
\end{equation}
and
\begin{equation}
\label{eqn:Eqn5a}
{v_{n}^\text{obs}}\left( {{\pt},\eta  > 0} \right)\  =\   \langle\langle\cos \left[ {n\left\{ { {\phi  - {\Psi _m}\left( {\mathrm{HF}-} \right)} } \right\}} \right] \rangle\rangle,
\end{equation}
where
\begin{equation*}
\langle\langle  f\left( \phi  \right) \rangle\rangle  = \frac{1}{(\text{\# Events})}\sum\limits_{j = 1}^\text{\# Events} \sum\limits_{i = 1}^{N_j} f\left(\phi_{i,j} \right).
\end{equation*}
In Eqs.~(\ref{eqn:Eqn5}) and (\ref{eqn:Eqn5a}) the first sum is over all particles $i$ with azimuthal angles $\phi_{i,j}$ within a given pseudorapidity  range in an event $j$ with
a given $\Psi_{m}$ and then a sum is taken over all events.
Particles with $\eta<0$ are correlated with HF$+$, and
those with $\eta>0$ are correlated with HF$-$.  In this manner, a minimum
gap of 3 units in pseudorapidity is maintained between the particles used in the
event plane angle determination and those used to determine the azimuthal anisotropy harmonic, helping in the suppression of short-range nonflow effects.

\begin{figure}[hbtp]
  \begin{center}
    \includegraphics[width=\cmsFigWidthStd]{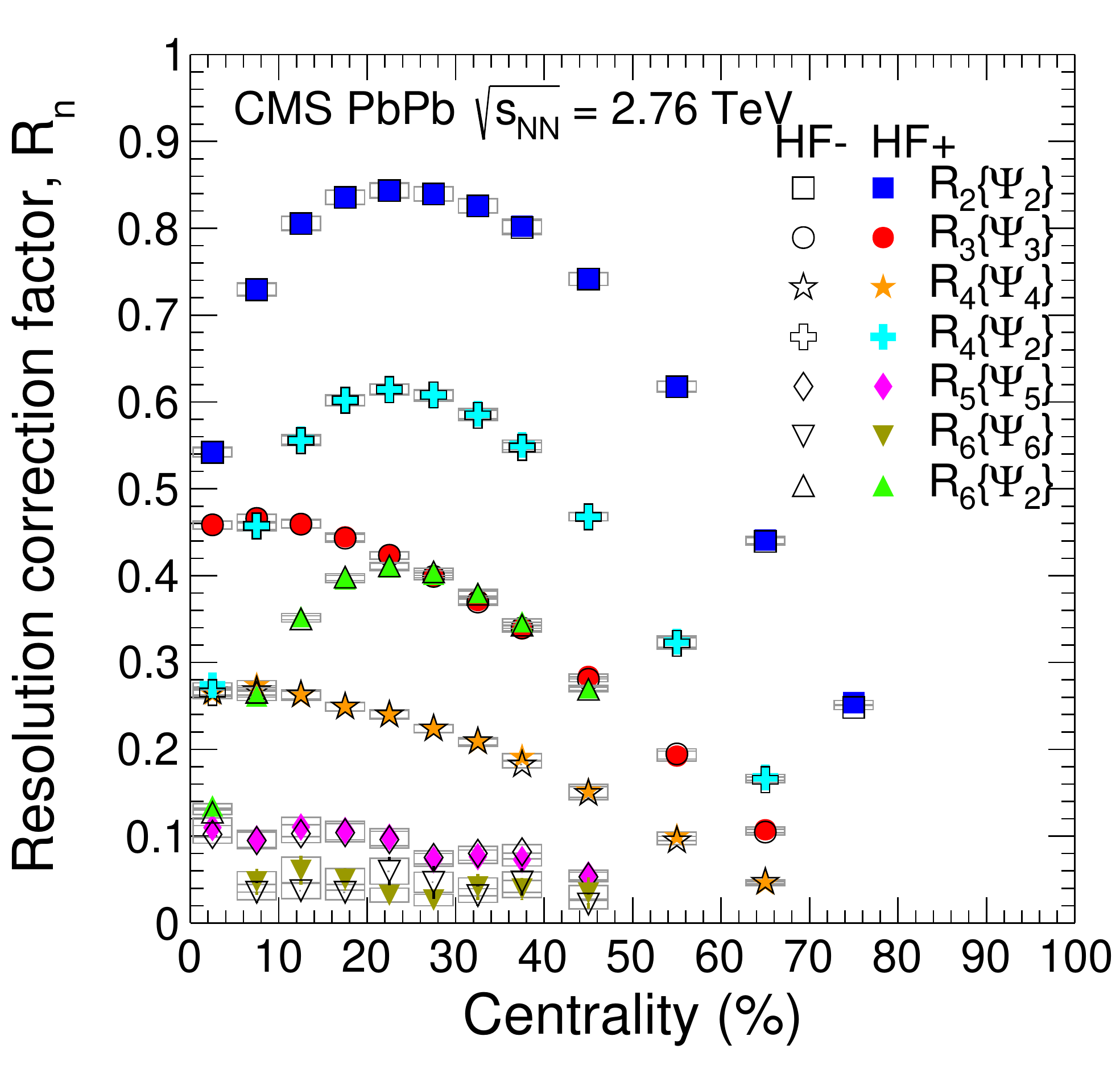}
    \caption{(Color online) Event plane resolution correction factors $R_n$ corresponding to
    different event plane angles $\Psi_m$ used in the analysis, as discussed in the text, are shown as a function of centrality for event planes determined with HF$-$ (open symbols) and HF$+$ (filled symbols).   The ${R}_2\{\Psi_2\}$ values are from Ref.~\cite{PhysRevC.87.014902} and are included for comparison purposes. Statistical uncertainties are smaller than the symbols.
The heights of the open gray rectangles indicate the systematic uncertainties.
 }
    \label{fig:Fig2}
  \end{center}
\end{figure}

The observed harmonic coefficient $v_{n}^\text{obs}$ depends on the resolution
of the event plane angles and is therefore sensitive to both the particle
multiplicity and the magnitude of the azimuthal asymmetry in the pseudorapidity range
used to determine the event plane angle.  The final anisotropy values are obtained by
correcting for the event plane angle resolution using a correction factor
$R_n\{\Psi_m\}$, with
$v_{n}\{\Psi_{m}\}=v_{n}^\text{obs}\{\Psi_m\}/R_{n}\{\Psi_{m}\}$.
To determine $R_n\{\Psi_m\}$ we use the three-subevent
method~\cite{PhysRevC.58.1671}. A subevent can be defined by restricting the particles used in an event plane determination to those found in a particular pseudorapidity range. The corresponding subevent angle is calculated only using particles in this limited region of pseudorapidity.   The  resolution
of $\Psi_{m}^{a}$  associated with subevent
$a$ (e.g., HF$-$) is determined using additional separate subevent
angles $\Psi_{m}^{b}$ (e.g., HF$+$) and  $\Psi_{m}^{c}$,
with
\begin{equation}
\label{eqn:Eqn6}
R_{n}^a\{\Psi_{m}\} = \sqrt {\frac{{\left\langle {\cos \left[ {n\left( {\Psi _m^a - \Psi _m^b} \right)} \right]} \right\rangle \left\langle {\cos \left[ {n\left( {\Psi _m^a - \Psi _m^c} \right)} \right]} \right\rangle }}{{\left\langle {\cos \left[ {n\left( {\Psi _m^b - \Psi _m^c} \right)} \right]} \right\rangle }}} .
\end{equation}
For this analysis $\Psi_m^c$ is determined using the silicon tracker detector.  Charged particles
with $\abs{\eta}<0.75$ are used to determine the  $\Psi_m^c$  subevent angle
based on  Eq.~(\ref{eqn:Eqn4}).  In this case the weight $w_i$ of particle $i$ is taken as the transverse momentum of the particle.  The resolution correction values used in the
analysis are shown in Fig.~\ref{fig:Fig2}. For the second-order event plane ($\Psi_{2}$; $m=2$) the resolution corrections are shown for the $n = 2$, 4, and 6 harmonic terms of Eq.~(\ref{eqn:Eqn3}). In general, the symmetry of the HF$+$ and HF$-$ detectors leads to very similar $R_n\{\Psi_m\}$ values for the event planes corresponding to the two detectors. However, when the resolution correction factors become sufficiently small, statistical fluctuations  start to have a significant influence on the observed values, resulting in differences between the HF$+$ and HF$-$ values.  In Fig.~\ref{fig:Fig2} this is evident for $R_n\{\Psi_m\}$ values less than 0.1. The contribution of the resolution correction to the overall $v_n$ systematic uncertainty is based on the observed difference in the HF$+$ and HF$-$ results.

At low \pt ($<0.8\GeVc$) the fraction of misreconstructed tracks is significant, reaching levels of up to 5\% at mid-rapidity and up to 25\% at forward rapidity ($\abs{\eta}\approx 2$) for the most central events. In this kinematic region, the $v_n$ signal is small. Full CMS Monte Carlo simulations based on the \textsc{hydjet}~\cite{EurPhysJC.45.211} and
\textsc{ampt}~\cite{PhysRevC.72.064901} event generators indicate that the $v_n$ signal
from misreconstructed tracks is approximately constant in this low-\pt region with a value that
can be larger than that of the properly
reconstructed tracks.  These studies suggest that the $v_n$ values from misreconstructed tracks can be characterized by a scaling factor $\kappa$, with
$v_n^{\text{mis}} = \kappa\left\langle {{v_n}} \right\rangle$, and where the
efficiency- and yield-weighted average $\left\langle {{v_n}} \right\rangle$
is performed over the  range
$0.3< \pt < 3\GeVc$.
Letting $f$ represent the proportion of misreconstructed tracks, the observed $v_n^\text{obs}$ value
is related to the ``true" flow signal $v_n^{\mathrm{real}}$ and that of the misreconstructed tracks
$v_n^{\mathrm{mis}}$
by
\begin{equation}
\label{eqn:Eqn2}
v_n^{\text{obs}}=(1-f)v_n^{\text{real}}+fv_n^{\text{mis}}.
\end{equation}
The scaling factor $\kappa$ is set by finding the value that leads to the least sensitivity
of the final results when varying the track criteria and hence the number of misreconstructed tracks, as determined by the Monte Carlo simulations.  The value is dependent on the harmonic and found to be
$\kappa=1.3\pm 0.1$ for
$v_2$,  $1.0\pm 0.4$ for $v_3$, and $0.8 \pm 0.6$ for all higher harmonics, where the
uncertainties are based on the observed sensitivity of the final results on the track criteria. The correction for misreconstructed tracks is negligible for \pt values above ${\approx}0.8\GeVc$.

\subsubsection{Cumulant method}
The cumulant method measures flow based on a cumulant expansion of multiparticle
azimuthal correlations. An ``integral" flow, or reference flow, of order $m$ is first determined using
an integral generating function of the multiparticle correlations
in a complex (imaginary) plane~\cite{PhysRevC.64.054901}.  The reference flow
plays a similar role to that of the
event plane determination in the event plane method.  The integral generating function is constructed using
all particles in a broad ($\pt$, $\eta$) window, averaging over the events in a
given centrality class.  Then, the differential flow, i.e., the flow in a narrower
phase space window, is measured with respect to the
reference flow. A particle in the differential bin is correlated to the particles
used for the reference flow through a differential generating function. Expressing the differential flow in terms of \pt bins, a reference flow of order $m$ can be used to determine differential flow of order $n$, where $n$ is an integral multiple of $m$.  To avoid
autocorrelations, if a given particle is used in determining the differential flow, the
particle will be excluded  in the calculation of the reference flow.
The generating functions for the reference and for differential
flows are calculated at several different points in the complex plane and then interpolated. The interpolation points are defined by
$z_{j,k}=r_0\sqrt  j~\text{exp}({i{2k\pi}/{8}})$, where $r_0\sqrt{j}$ is the radius of a point
and ${2k\pi}/{8}$ its polar angle.
In this analysis, the counting indices are taken as $j = 1$, 2, 3 and $k = 0$--7.
Two (three) values for the radius parameter $r_0$ are used for the differential (reference)
flow.  The radius parameters are determined according to the detected
charged particle multiplicity and the number of events analyzed in each centrality class.
To reduce the effect of  fluctuations in the event-by-event multiplicity on the flow measurements,  for each given
centrality interval, 80\% of the mean multiplicity
$\langle M\rangle$  of particles for that interval are selected at random in each event to
determine the reference flow. In addition, the transverse momentum is restricted to
$\pt <3$\GeVc in the reference flow to reduce the nonflow contribution arising mainly from jets. The cumulant $v_{3}$ coefficient
is measured with four particle correlations and denoted
$v_{3}\{4\}$, with the differential and integral flow both of order $m=n=3$. The use of a four particle correlation strongly reduces the influence of the nonflow effects that are evident in
two particle cumulant analyses~\cite{PhysRevC.87.014902}. The cumulant  $v_{4}$ coefficient is calculated relative to the integral $v_{2}$ behavior
using a five particle correlation and
is denoted  $v_{4}\{5\}$.
It is not possible to extract flow coefficients with $n>4$ using this method since the signals become too small.
However, $v_{6}$ coefficients calculated relative to the integral $v_{2}$ behavior are obtained using the Lee--Yang zeros method, as discussed below.

If the detector had uniform acceptance and full efficiency, a reference flow value calculated for a given
pseudorapidity range would be equivalent to the yield-weighted flow coefficient for the same phase space. The need for an efficiency correction for the differential-flow coefficients is avoided by choosing sufficiently small \pt bins such that the efficiency does not change significantly across the bin. However, this is not true for the reference flow and, to account for efficiency and acceptance effects, the yield-weighted cumulant $v_{n}$ values presented in this paper are based on the corresponding differential flow coefficient $v_{n}(\pt)$ weighted by the efficiency and acceptance corrected yields, similar to what is done for the event plane analysis.
For both the multiparticle cumulant method and the Lee--Yang zeros method discussed in the next subsection, the influence of misreconstructed tracks at low-\pt values is found to have only a small effect on the final results and is included as part of the
systematic uncertainty.

\subsubsection{Lee--Yang zeros method}
The Lee--Yang zeros (LYZ) method~\cite{NuclPhysA.727.373, Borghini:2004ke} studies directly the
large-order behavior of the cumulant expansion of the azimuthal correlations. As done for the cumulant analysis, a reference flow of order $m$ is first determined based on a complex generating function that is calculated over a large range of momentum and pseudorapidity for a given centrality range. The generating function can be expressed in terms of either a sum or a product of individual particle terms. The product form is used in this analysis since it is expected to be less sensitive to nonflow and autocorrelation effects~\cite{Borghini:2004ke}. An estimate of the reference flow is found in terms of the location of the first minimum of the generating function calculated for a fixed projection angle.  The analysis uses five different projection angles and averages the results to reduce statistical uncertainties.
Charged particles with $0.3 < \pt < 12\GeVc$ and $ \abs{\eta}<2.4 $ are used to calculate the
reference flow, which corresponds to the yield-weighted average flow in the indicated phase space, but
neglects efficiency and acceptance effects.

Once the reference flow has been established, the differential flow $v_{n}\{\mathrm{LYZ}\}(\pt)$ in a limited \pt and pseudorapidity range can be determined with respect to the generating function of the reference flow.  Again, the differential flow harmonic $n$ can be any integral multiple of the reference flow harmonic $m$. As with the cumulant analysis, the integral $v_{n}\{\mathrm{LYZ}\}$ coefficients presented in this paper are obtained by taking a yield-weighted average of the differential flow results, after correcting the yield for efficiency and acceptance effects.  The new CMS Lee--Yang zeros results are for the $n = 4$ and 6 harmonics based on the $m = 2$ order reference flow. Measurements of the $n = 3$ and 5 harmonics are not possible because of their small magnitudes. Details of this part of the analysis can be found in Ref.~\cite{PhysRevC.87.014902}.

\subsection{Systematic uncertainties}
The systematic uncertainties include those common to all methods, as well as
method specific ones. Tables of uncertainties for the results presented in this paper are given in the Appendix. Common uncertainties
include those resulting from the tracking efficiency and the centrality determination.
Protons, pions, and kaons can have different $v_n$ values, and therefore species dependent
differences in their tracking
efficiency will affect the unidentified, charged particle results. The influence of particle composition is studied by determining the efficiency for identified particle detection using simulations of the CMS detector and then making different assumptions
of the \pt dependence of the particle mix based on the \textsc{hydjet} event generator~\cite{EurPhysJC.45.211} and on assuming a similar behavior as found at RHIC~\cite{PHENIX_pid}.
As previously done to account for this
effect in the $v_2$ measurement~\cite{PhysRevC.87.014902}, a
conservative 0.5\% uncertainty, independent of $\pt$, $\eta$, and centrality, is assumed.
The sensitivity of the harmonic-flow coefficients to
the centrality calibration is evaluated by varying the trigger efficiency by $\pm3$\%.
The resulting uncertainty in $v_n$ is of the order of 1\% and this value is adopted independent of
\pt and centrality. The uncertainty in the overall charged particle efficiency corrections, which only
affects the yield-weighted average $v_{n}$ values,  is evaluated by
determining the efficiency based on the \textsc{hydjet} model and, separately, by embedding
simulated pions into recorded PbPb collision event data. The difference between the two resulting efficiencies gives at most a 0.5\% change in the yield-weighted average $v_n$ values, which is taken as a systematic
uncertainty.

Misreconstructed tracks affect both the
differential $v_n(\pt)$ results and the yield-weighted average $v_{n}$ measurements of all methods,
although not necessarily to
the same extent. Therefore, separate studies are performed for each method.  The effect of misreconstructed tracks, evaluated by varying the track quality criteria and labeled as the ``Track Quality Requirements" uncertainty in the Tables~\ref{tab:sys_v3_EP_combined} to~\ref{tab:sys_v6_LYZ_combined},
generally accounts for the largest single contribution to the systematic uncertainty, especially at
low \pt and for the most central events. Different sets of track quality requirements on the pixel tracks
are used, and the ratio of the results provide an estimate of the systematic uncertainty
from this source in different \pt ranges. Track quality requirements include having the track pointing back to within a specified range of the reconstructed vertex location and having a specified goodness-of-fit for the track.

For the event plane method, the uncertainty in the resolution correction value is  primarily a consequence of its
statistical uncertainty and is generally
small compared to the track quality requirement uncertainty.  This is seen in Tables~\ref{tab:sys_v3_EP_combined} to~\ref{tab:sys_v6psi2_EP_combined}
where the systematic uncertainties for the $v_{n}(\pt)$
values obtained using the event plane method are presented. However, for the $v_5$ and $v_6$ results the resolution correction uncertainty becomes comparable to that for the track quality requirement uncertainty for mid-central events. The various systematic uncertainties are taken to be
uncorrelated and added in quadrature in the measurements of the $v_n$ coefficients.

In addition to the systematic uncertainty terms common to all methods,
the cumulant analyses are also influenced by the choice of the $r_0$ radius parameter and by the effect of
fluctuations in the event-by-event multiplicity on the reference flow.
These uncertainties are estimated by varying the $r_0$ parameter and comparing the flow results
with and without the selection of 80\% of the mean multiplicity.
Tables~\ref{tab:sys_v3_4part_combined} and \ref{tab:sys_v4_5part_combined}
show the systematic uncertainties associated with the $v_{3}\{4\}$ and $v_{4}\{5\}$ results. The \pt dependence of
$v_{3}\{4\}$ could only be measured up to 4\GeVc because of the small amplitude of this coefficient.

The effect of fluctuations in the event-by-event multiplicity is
studied for the LYZ method by analyzing the events in finer 2.5\% centrality bins.
Tables~\ref{tab:sys_v4_LYZ_combined} and \ref{tab:sys_v6_LYZ_combined} show
the systematic uncertainties in the $v_{4}\{\mathrm{LYZ}\}$ and $v_{6}\{\mathrm{LYZ}\}$ results, respectively.
In these cases, the
total uncertainties, again found by adding the component uncertainties in quadrature, are dominated by
the track quality requirement uncertainties.

The systematic uncertainties for the yield-weighted, average $v_{n}$ values with $0.3< \pt < 3.0\GeVc$ are also shown in Tables~\ref{tab:sys_v3_EP_combined} to~\ref{tab:sys_v6_LYZ_combined}. This \pt range is the same as used in the previous CMS elliptic-flow
analysis~\cite{PhysRevC.87.014902} and covers the \pt range dominated by hydrodynamic flow.

\subsection{Glauber model eccentricity calculations}
\label{sec:glauber}
The Glauber model treats a nucleus-nucleus
collision as a sequence of independent nucleon-nucleon collisions (see Ref.~\cite{AnnRevNuclPart.57.205}
and references therein). The model can be used to obtain anisotropy parameters based on the transverse location of
the participant nucleons~\cite{Alver:2008aq}. These, in turn, are
expected to be reflected in the observed particle anisotropies.
The Glauber model assumes that the nucleons
in a nucleus are distributed according to a Woods-Saxon density
\begin{equation}
\rho(r) = \frac{\rho_{0}(1+wr^2/R^2)}{1+\re^{(r-R)/a}},
\end{equation}
where $\rho_{0}$ is the nucleon density in the center of the nucleus,
$R$ is the nuclear radius, $a$ is the skin depth, and $w$ characterizes
deviations from a spherical shape. For $^{208}$Pb, the parameters
$R = 6.62\pm 0.13$\unit{fm}, $a = 0.546\pm 0.055$\unit{fm}, and $w = 0$ are used~\cite{DeJager:1987qc}.
The model assumes that nucleons in each nucleus travel on straight-line trajectories through
the colliding system and interact according to the inelastic nucleon-nucleon cross section,
$\sigma_{\text{inel}}^{NN}$, as measured in pp collisions.
A value of $\sigma_{\text{inel}}^{NN} = 64 \pm 5$\unit{mb}, which is found through an interpolation of values obtained at
different center-of-mass energies~\cite{Beringer:1900zz}, is used in the calculations
at $\sNN= 2.76\TeV$.

The spatial anisotropies of order $n$ based on a participant plane angle of order $m$ are calculated using the transverse location of each participant~\cite{PhysRevC.82.064903},  with
\begin{equation}
\label{eqn:Eqn8}
{\epsilon _{n,m}} = \frac{\sqrt{{{\left\langle {r_ \bot ^n\cos \left[ {n\left( {\phi  - {\Phi _m}} \right)} \right]} \right\rangle}^2 }}}{{\left\langle {r_ \bot ^n} \right\rangle }},
\end{equation}
where, for a participant located at coordinates $\{x,y\}$ in the transverse plane,
$r_ \bot ^n = {\left( {\sqrt {{x^2} + {y^2}} } \right)^n}$, and $\phi  = \arctan \left( {y/x} \right)$.  The ``participant plane" angle $\Phi_{m}$ is then found by summing over all participant particles with
\begin{equation}
\label{eqn:Eqn9}
{\Phi _m} = \frac{1}{m}{\tan ^{ - 1}}\left\{ {\frac{{\left\langle {r_ \bot ^m\sin \left[ {m\phi } \right]} \right\rangle }}{{\left\langle {r_ \bot ^m\cos \left[ {m\phi } \right]} \right\rangle }}} \right\}.
\end{equation}
For $n=m$ we define $\epsilon_{n} = \epsilon_{n,n}$.
With this definition, $\epsilon_{n}$ (or $\epsilon_{n,m}$) can only
take positive values and represents the maximum asymmetry for each collision. It has been demonstrated that a
common behavior is achieved for the elliptic-flow coefficient $v_2$ in AuAu and CuCu collisions at
$\sNN=200$\GeV when scaled by the eccentricity $\epsilon_{n}$ and plotted as a function
of the transverse charged particle areal density~\cite{PhysRevC.77.014906}.  This scaling persist at
LHC energies~\cite{PhysRevC.87.014902}.

Table~\ref{ecc_rn} lists the results for the average number of
participants, $\langle N_\text{part} \rangle$, and the root-mean-squared
evaluation of the eccentricities,
$\sqrt{\langle\epsilon_{n}^{2}\rangle}$ or $\sqrt{\langle\epsilon_{n,m}^{2}\rangle}$,  and the corresponding
systematic uncertainties for the centrality bins used in this analysis. The method used to convert from impact parameter to centrality is discussed
in Ref.~\cite{PhysRevC.84.024906}. The uncertainties in the parameters involved in the Glauber model calculations contribute to the systematic uncertainty in ${N_\text{part}}$ and $\epsilon_n$ for a given centrality bin.

\begin{table*}[ht]
\ifthenelse{\boolean{cms@external}}{}{\tiny}
\centering
\topcaption{The average number of participating nucleons and participant eccentricities, weighted by $r^n$, calculated using the Glauber model in bins of centrality.  Systematic uncertainties resulting from the
uncertainties in the Glauber-model parameters are indicated.}
\label{ecc_rn}
\resizebox{\textwidth}{!}{
\newcolumntype{x}[1]{D{+}{\pm}{#1,#1}}
\begin{scotch}{cx{3}x{3}x{3}x{3}x{3}x{3}x{3}x{3}}
\textbf{Centrality} &\multicolumn{1}{c}{${\langle N_\text{part}\rangle}$ }&\multicolumn{1}{c}{${ \sqrt{\langle\smash[b]{\epsilon_{2}^{2}}\rangle}}$ }&\multicolumn{1}{c}{${ \sqrt{\langle\smash[b]{\epsilon_{3}^{2}}\rangle}}$}&\multicolumn{1}{c}{${ \sqrt{\langle\smash[b]{\epsilon_{4}^{2}}\rangle}}$ }&\multicolumn{1}{c}{${ \sqrt{\langle\epsilon_{5}^{2}\rangle}}$}&\multicolumn{1}{c}{${ \sqrt{\langle\smash[b]{\epsilon_{6}^{2}}\rangle}}$} &\multicolumn{1}{c}{${ \sqrt{\langle\smash[b]{\epsilon_{4,2}^{2}}\rangle}}$}&\multicolumn{1}{c}{${ \sqrt{\langle\smash[b]{\epsilon_{6,2}^{2}}\rangle}}$} \\
\textbf{(\%)} &&&&&&&& \\
\hline
0--5 &  381+2 &0.084+0.010 &0.097+0.010 &0.114+0.010 &0.131+0.010 &0.149+0.010 &0.081+0.041 &0.106+0.065 \\
5--10 &  329+3 &0.127+0.010 &0.129+0.010 &0.148+0.010 &0.169+0.010 &0.190+0.010 &0.104+0.064 &0.134+0.081 \\
10--15 &  283+3 &0.175+0.011 &0.154+0.010 &0.174+0.010 &0.198+0.010 &0.220+0.010 &0.123+0.059 &0.156+0.092 \\
15--20 &  240+3 &0.219+0.016 &0.177+0.010 &0.199+0.010 &0.225+0.010 &0.248+0.011 &0.143+0.049 &0.176+0.081 \\
20--25 &  204+3 &0.262+0.016 &0.199+0.010 &0.225+0.010 &0.250+0.010 &0.274+0.013 &0.165+0.049 &0.194+0.073 \\
25--30 &  171+3 &0.301+0.019 &0.221+0.010 &0.254+0.010 &0.277+0.010 &0.302+0.014 &0.193+0.038 &0.213+0.062 \\
30--35 &  143+3 &0.339+0.022 &0.245+0.010 &0.284+0.011 &0.307+0.011 &0.331+0.015 &0.221+0.039 &0.235+0.062 \\
35--40 &  118+3 &0.375+0.022 &0.268+0.011 &0.317+0.013 &0.337+0.012 &0.361+0.015 &0.254+0.041 &0.257+0.067 \\
40--50 & 86.2+2.8 &0.429+0.024 &0.308+0.013 &0.370+0.016 &0.385+0.016 &0.410+0.017 &0.307+0.035 &0.297+0.070 \\
50--60 & 53.5+2.5 &0.501+0.026 &0.366+0.015 &0.445+0.020 &0.454+0.018 &0.475+0.018 &0.385+0.039 &0.355+0.075 \\
60--70 & 30.5+1.8 &0.581+0.027 &0.422+0.016 &0.520+0.023 &0.513+0.018 &0.534+0.020 &0.466+0.039 &0.417+0.069 \\
70--80 & 15.7+1.1 &0.662+0.026 &0.460+0.012 &0.596+0.026 &0.559+0.015 &0.609+0.023 &0.549+0.035 &0.497+0.063 \\
\end{scotch}
}
\end{table*}

\section{Results}
\label{sec:Results}
\label{sec:results}

This section presents the results for the higher-order harmonic coefficients. Previously published
two particle correlation results~\cite{Eur.Phys.J.C72.2012} from the CMS Collaboration are also presented for completeness.  The \pt dependence of the coefficients at mid-rapidity is presented first,
comparing the values obtained using the different analysis methods.   This is followed by the measurements of the yield-weighted average $v_{n}$ values,
which are given in terms of both their centrality and pseudorapidity dependencies.   We conclude the section with comparisons of
the CMS measurements to published results of the ALICE~\cite{PhysRevLett.107.032301,PhysLettB.708.249} and ATLAS~\cite{PhysRevC.86.014907} Collaborations.

\subsection{\texorpdfstring{The \pt dependence of $v_n$ at mid-rapidity}{The pt dependence of vn at mid-rapidity}}

\begin{figure*}[hbtp]
  \begin{center}
    \includegraphics[width=\cmsFigWidthStdWide]{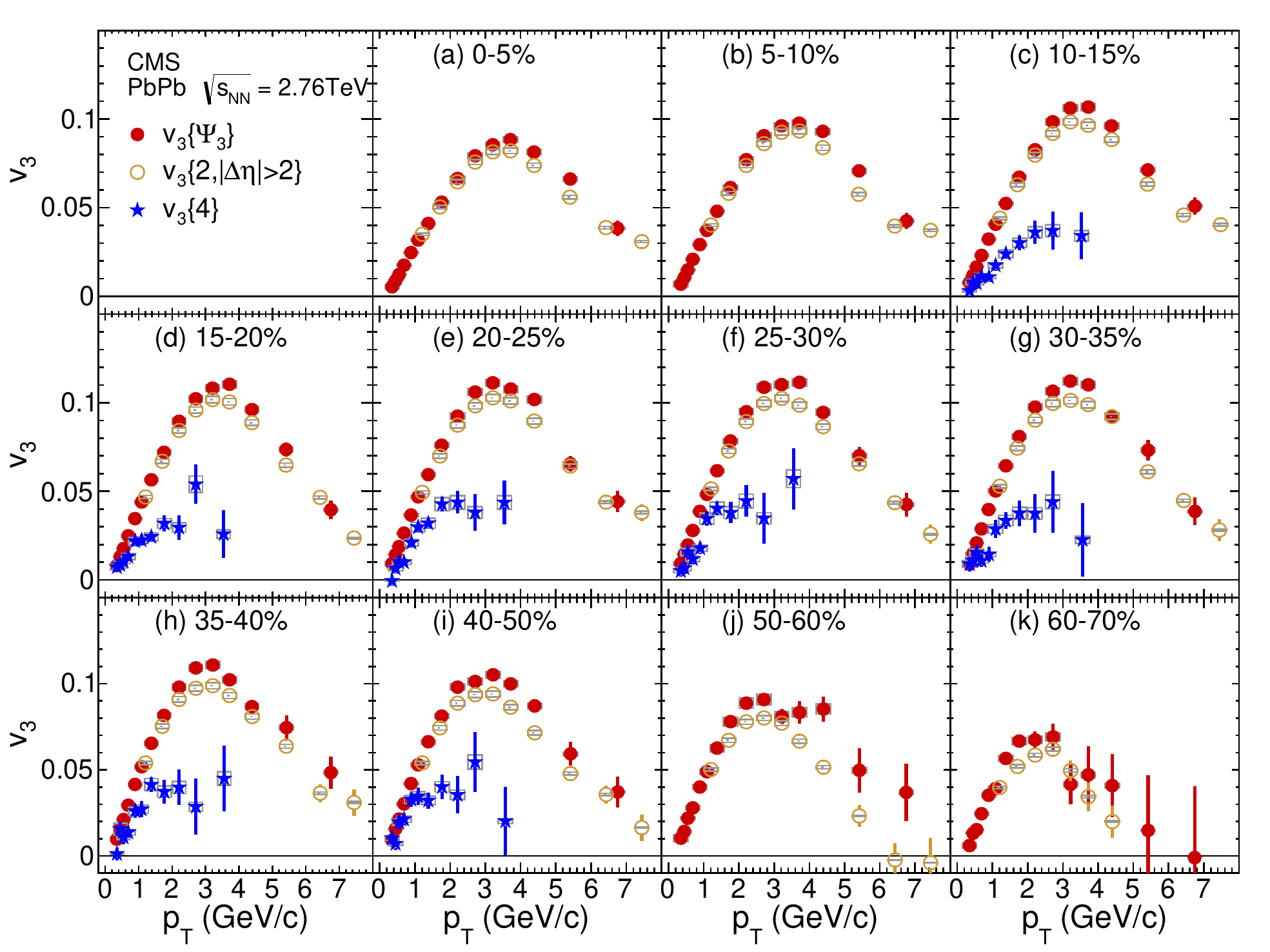}
    \caption{(Color online) Measurements of the azimuthal asymmetry coefficient $v_{3}$ from
    three different methods as a function of \pt for the indicated centrality bins, as specified in percent. The event plane (filled circles) and cumulant (filled stars) results are with  $\abs{\eta} < 0.8$. The two particle correlation results (open circles) are from a previous CMS measurement~\cite{Eur.Phys.J.C72.2012}.  Statistical (error bars) and systematic (light gray boxes) uncertainties are shown.}
    \label{fig:Fig4}
  \end{center}
\end{figure*}

Figure~\ref{fig:Fig4} shows the $v_{3}$ coefficient results
for $\abs{\eta}<0.8$ based on the event plane $v_{3}\{\Psi_{3}\}$ and four particle cumulant $v_{3}\{4\}$
methods. The analyses employ the same event selection as used in the previous elliptic-flow ($n=2$) study of Ref.~\cite{PhysRevC.87.014902}. Also shown are the two particle correlation results
$v_{3}\{{2,\abs{\Delta\eta}>2}\}$ of  Ref.~\cite{Eur.Phys.J.C72.2012}.
The two particle correlation method is similar to a two particle cumulant analysis, as
used in Ref.~\cite{PhysRevC.87.014902}, but requires a pseudorapidity gap between the two particles, which was not the case for the two particle cumulant analysis.  For the two particle correlation method, charged particles with
$\abs{\eta}<2.5$ are first selected. To reduce the effect of nonflow processes, particle pairs are chosen with the requirement of a pseudorapidity gap between the particles in each pair of $2<\abs{\Delta\eta}<4$. This method,
as applied to LHC data, is described in detail in
Refs.~\cite{PhysLettB.708.249,Eur.Phys.J.C72.2012}.

The event plane and two particle correlation results are found to be very similar, although the results from the two particle correlations have systematically smaller values.   This suggests similar sensitivity to initial state fluctuations and  nonflow effects for the current implementations of the two methods when a large pseudorapidity gap
is required for both analyses. We also note that the values of the harmonic coefficients determined from the
two particle correlation  method  correspond to
$\sqrt{\langle v_{n}^2\rangle}$.  For near-perfect event plane resolution with $R_{n}\approx 1$ (Eq.~(\ref{eqn:Eqn6})), the event plane results are expected to approach $\langle v_{n}\rangle$, whereas for lower values of $R_{n}$, the event plane method gives values closer to $\sqrt{\langle v_{n}^2\rangle}$~\cite{PhysRevC.77.014906}. We discuss this later in the paper.  With the $R_{n>2}$ values shown in Fig.~\ref{fig:Fig2}, the event plane method is expected to produce values very close to $\sqrt{\langle v_{n}^2\rangle}$ in the absence of fluctuations.

The smaller values found for the two particle correlation method is likely a consequence of sampling a larger pseudorapidity range than the event plane analysis, which only considered particles with $\abs{\eta}<0.8$. As shown later in this paper, the spectrum-weighted harmonic coefficients are found to have their maximum values near $\eta\approx 0$ (see Section~\ref{IntVN}). The pseudorapidity gap used in the two particle correlations prevents the selection of both particles in a pair from the mid-rapidity, maximum $v_3$ region, thus assuring a somewhat smaller $v_3$ result as compared to the event plane analysis.
The significantly smaller $v_3\{4\}$ values for the
four particle cumulant method can be attributed to the difference in how fluctuations affect two particle and higher-order correlations~\cite{PhysRevC.80.014904}. Assuming a smooth overlap region and in the absence of fluctuations, the $v_3$ harmonic is expected to vanish based on the symmetry of the overlap region. For the two particle correlation and event plane results the fluctuations are expected to increase the harmonic coefficients with respect to those expected without initial state fluctuations, whereas for fourth- (and higher-) order particle correlations, the fluctuations can lower the values~\cite{PhysRevC.80.014904}. It was shown in Ref.~\cite{PhysRevC.87.014902} that effects of fluctuations can largely account for the differences observed in the $v_2$ values for the different methods. Fluctuations are expected to dominate the initial state geometry of the odd-harmonic, $n=3$ asymmetry, as discussed in Refs.~\cite{PhysRevC.81.054905,PhysRevC.82.064903,PhysRevC.84.034910,PhysRevC.84.024911}.

\begin{figure*}[hbtp]
  \begin{center}
    \includegraphics[width=\cmsFigWidthStdWide]{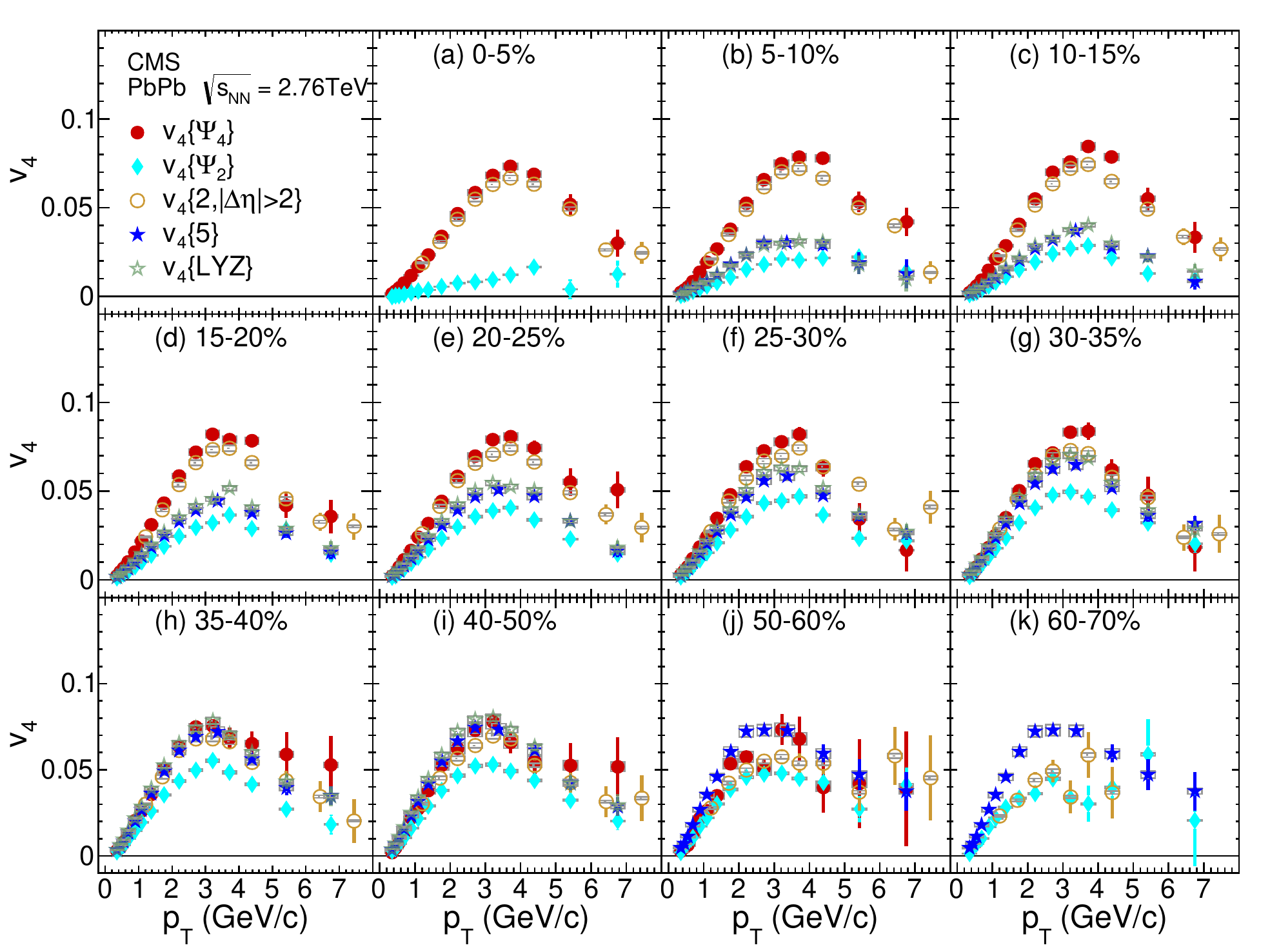}
    \caption{(Color online) Measurements of the azimuthal asymmetry coefficient $v_{4}$  from four different methods as a function of \pt for the indicated centrality bins, as specified in percent. The event plane (filled circles and filled diamonds), cumulant (filled stars), and Lee--Yang zeros (open stars) analyses are with
    $\abs{\eta} < 0.8$. The two particle correlation results (open circles) are from a previous CMS measurement~\cite{Eur.Phys.J.C72.2012}. Statistical (error bars) and systematic (light gray boxes) uncertainties are shown.}
    \label{fig:Fig5}
  \end{center}
\end{figure*}

\begin{figure*}[hbtp]
  \begin{center}
    \includegraphics[width=\cmsFigWidthStdWide]{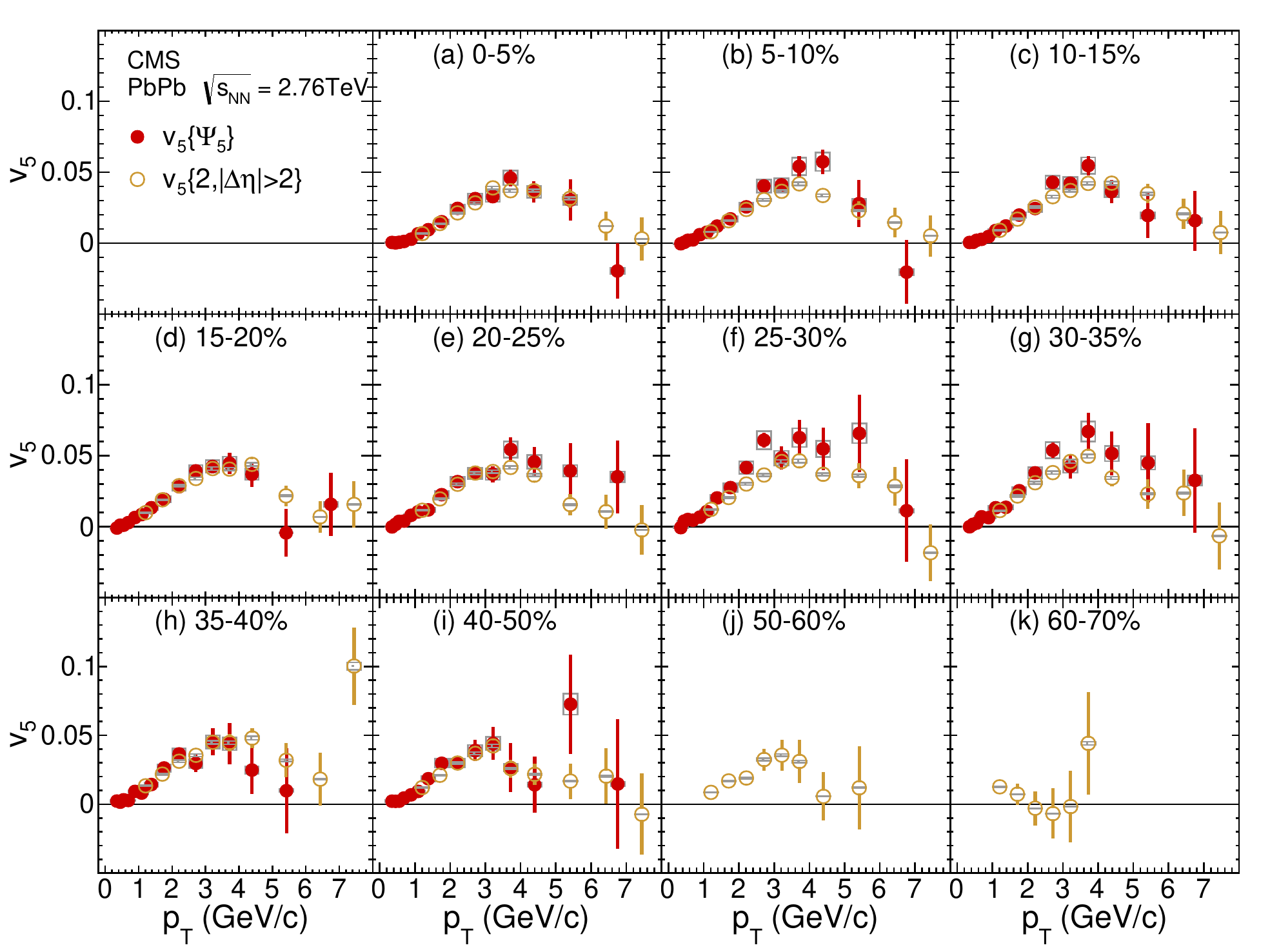}
    \caption{(Color online) Measurements of the azimuthal asymmetry coefficient $v_{5}$ from two different methods as a function of \pt for the indicated centrality bins, as specified in percent. The event plane analysis (filled circles) is with
    $\abs{\eta} < 0.8$. The two particle correlation results (open circles) are from a previous CMS measurement~\cite{Eur.Phys.J.C72.2012}. Statistical (error bars) and systematic (light gray boxes) uncertainties are shown.}
    \label{fig:Fig6}
  \end{center}
\end{figure*}

\begin{figure*}[hbtp]
  \begin{center}
    \includegraphics[width=\cmsFigWidthStdWide]{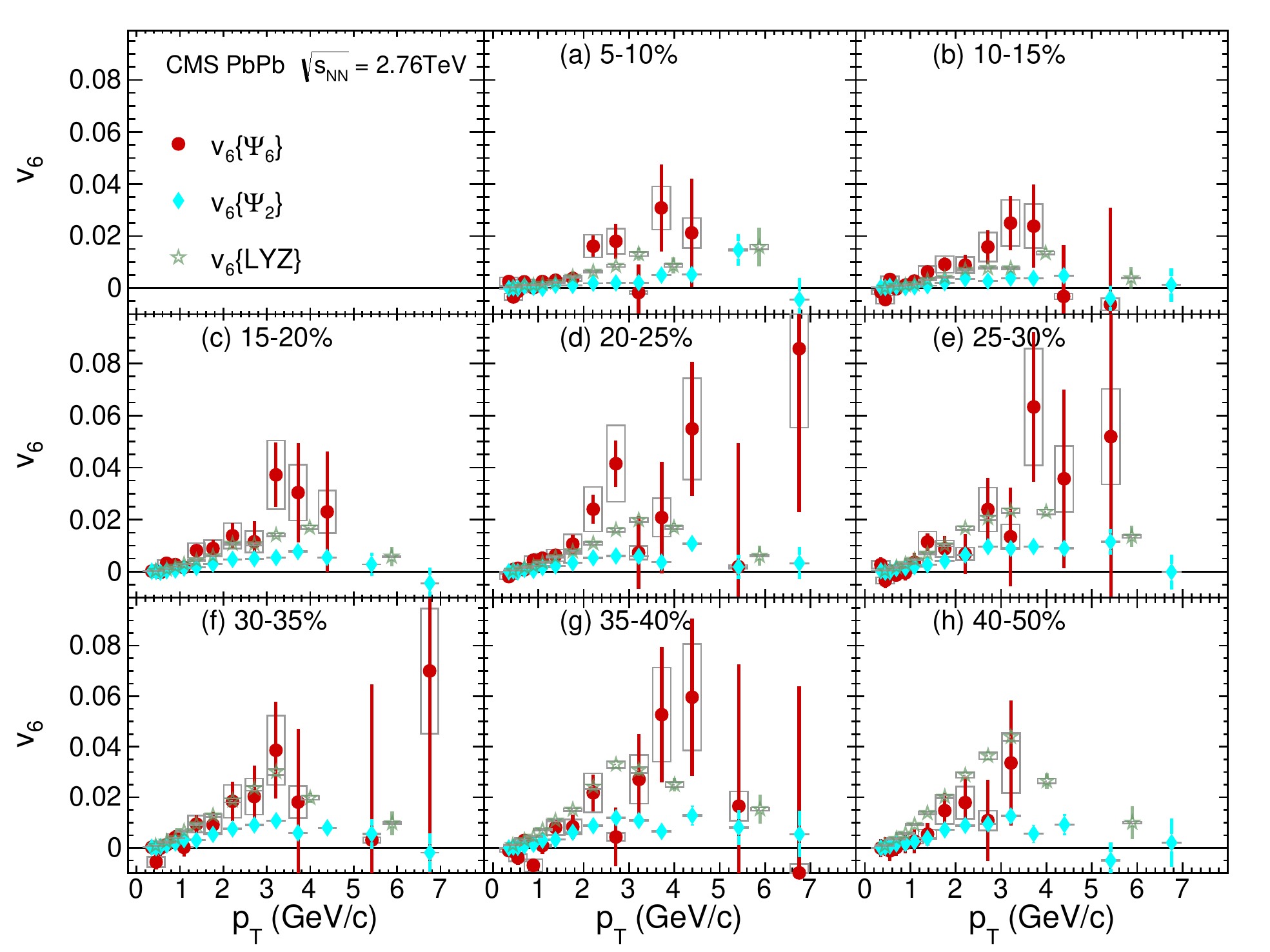}
    \caption{(Color online) Measurements of the azimuthal asymmetry coefficient $v_{6}$ from the event plane (filled circles and filled diamonds) and Lee--Yang zeros (open stars)  methods as a function of \pt for the indicated centrality bins, as specified in percent.  The results are for $\abs{\eta} < 0.8$. Statistical (error bars) and systematic (light gray boxes) uncertainties are shown.}
    \label{fig:Fig7}
  \end{center}
\end{figure*}

Figure~\ref{fig:Fig5} shows the $v_{4}$ values for a number of different methods.   The event plane
results are shown based on both the second-order, elliptic-flow event plane ($m=2$) as well as the
fourth-order ($m=4$) event plane.  A significant centrality dependence is observed for the $v_{4}\{\Psi_{2}\}$, $v_{4}\{5\}$, and $v_{4}\{\mathrm{LYZ}\}$ results, which are all based on a second-order reference distribution, while only a weak centrality dependence is found for the $v_{4}\{\Psi_{4}\}$ and $v_{4}\{{2,\abs{\Delta\eta}>2}\}$ values, these last two depending on a fourth-order reference distribution.

Figure~\ref{fig:Fig6} shows the $v_{5}$ anisotropy coefficients based on the event plane of
the same $m=5$ order and the two particle correlation method.   Similar to the other multipoles,
the two methods give very similar results, with only a small dependence on centrality for most of the range.

Finally, Fig.~\ref{fig:Fig7} shows the event plane results for $v_{6}$ based on both the $m=2$ and 6 event planes, as well as the LYZ results  based on the $m=2$ integral reference flow.  In this case, the event plane results based on the second-order reference distribution are consistently smaller than those found for either the LYZ method, which are also based on a second-order integral-flow behavior, or the event plane method using a sixth-order reference distribution.  The higher values and relatively weak centrality dependence found for the $v_{6}\{\Psi_{6}\}$ results are consistent with these values being strongly influenced by fluctuations.

Summarizing the results for this section, the differential azimuthal harmonics are found to have their strongest dependence on
centrality when the ``reference" particles are based on the second-order participant planes,
as is the case for the $v_{2}\{\Psi_{2}\}$,
$v_{2}\{{2,\abs{\Delta\eta}>2}\}$, $v_{2}\{4\}$, $v_{2}\{\mathrm{LYZ}\}$, $v_{4}\{\Psi_{2}\}$,  $v_{4}\{5\}$, $v_{4}\{\mathrm{LYZ}\}$, and $v_{6}\{\Psi_{2}\}$ results.  A weaker centrality dependence is observed in the other cases where the higher-order ($m>2$) reference plane is of the same order as the harmonic being studied.
This weak centrality dependence suggests a reduced influence of the average geometry of the
overlap region, as might be expected if fluctuations in the participant locations dominate the results.

\subsection{Yield-weighted average anisotropies}
\label{IntVN}

\begin{figure*}[hbtp]
  \begin{center}
    \includegraphics[width=\cmsFigWidthStdWide]{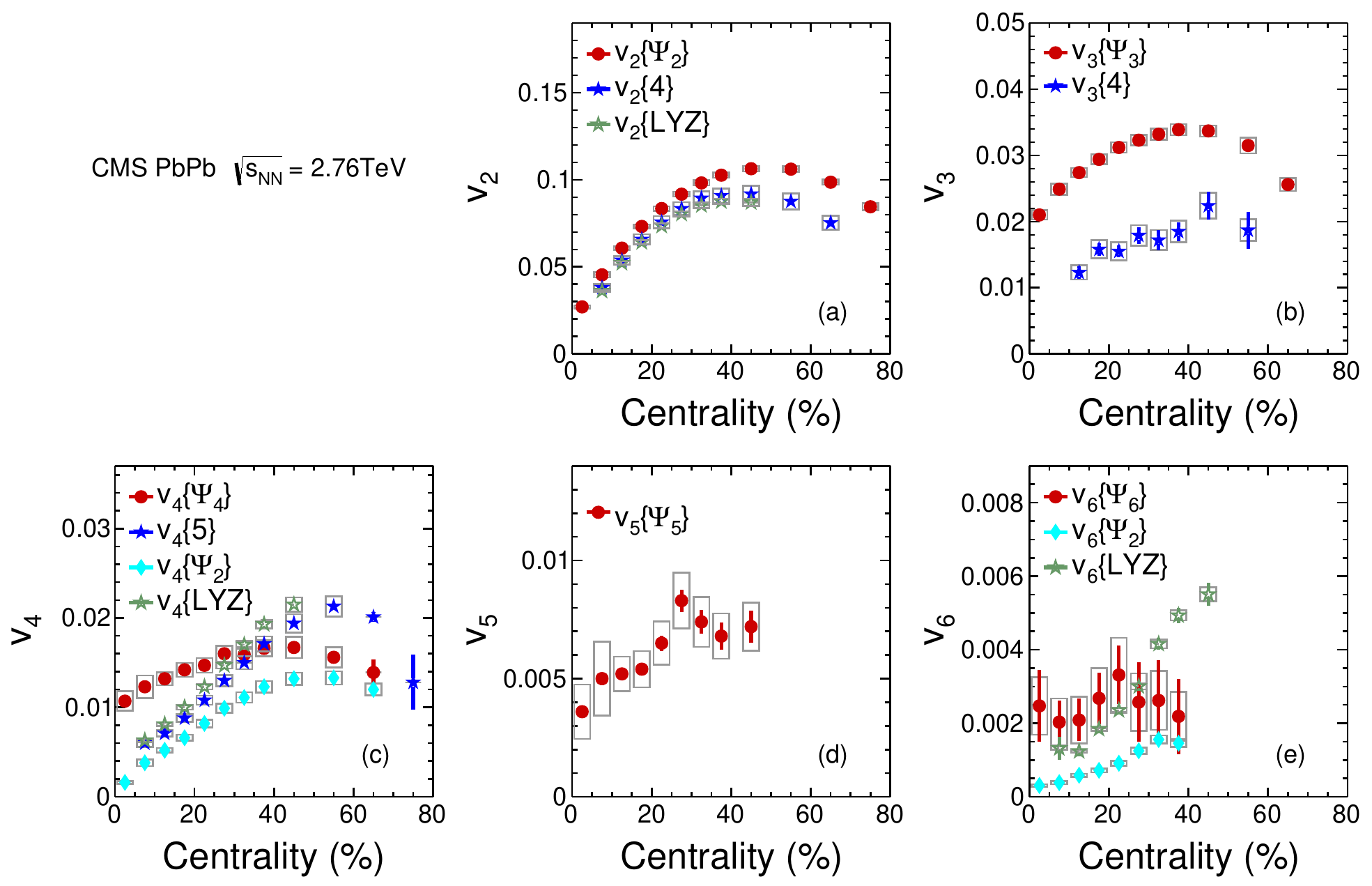}
    \caption{(Color online) Yield-weighted average azimuthal asymmetry coefficients $v_{n}$, for $n =2$--6, with $0.3<\pt<3.0\GeVc$ are shown for three different methods as a function of centrality. The $v_2$ results are from Ref.~\cite{PhysRevC.87.014902} and included for completeness. Statistical (error bars) and systematic (light gray boxes) uncertainties are shown. The different results found for a given $v_{n}$ reflect the role of participant fluctuations and the variable sensitivity to them in each method, as discussed in the text.}
    \label{fig:Fig8}
  \end{center}
\end{figure*}

\begin{figure*}[hbtp]
  \begin{center}
    \includegraphics[width=\cmsFigWidthStdWide]{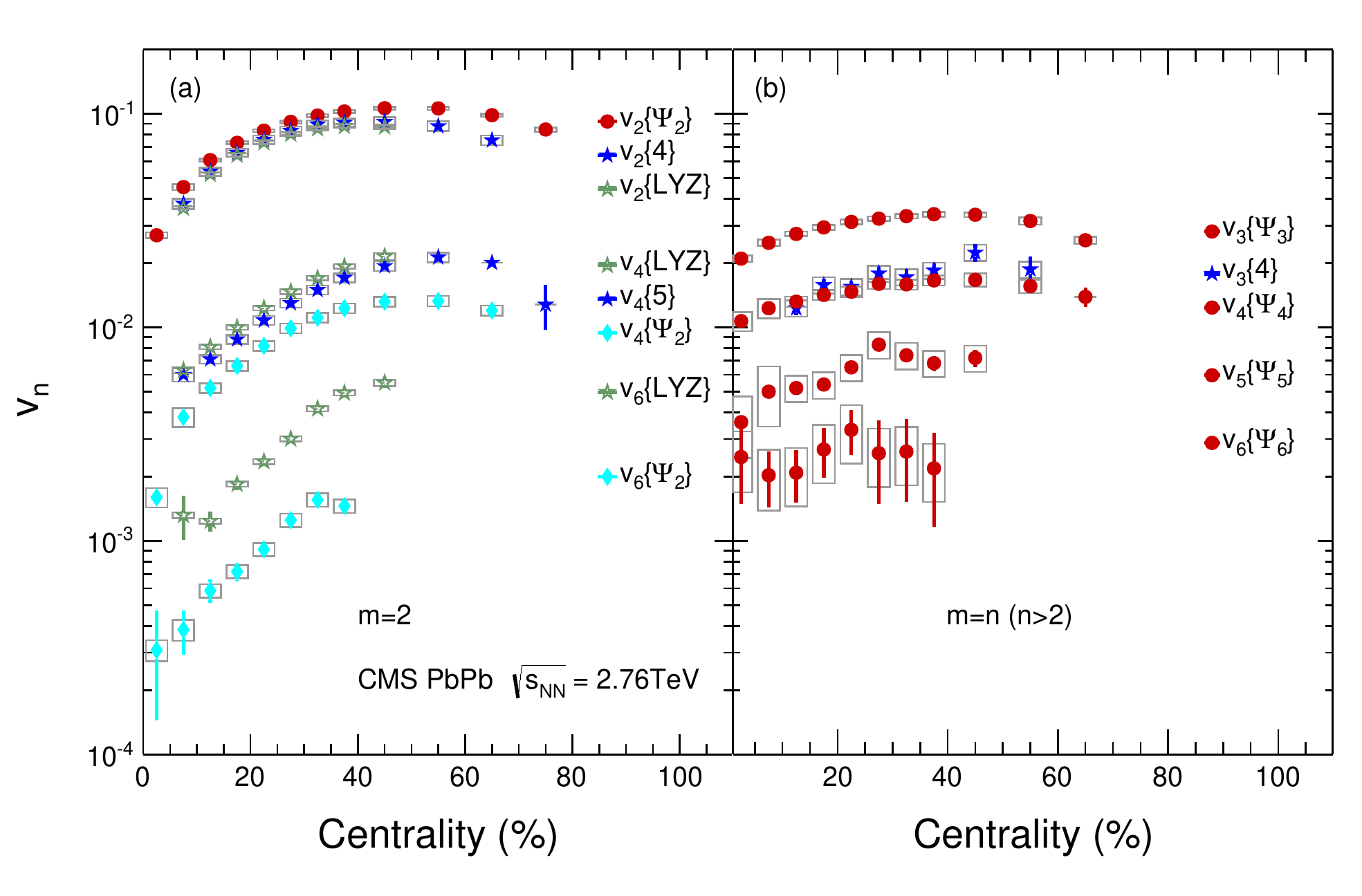}
    \caption{(Color online) (a) Yield-weighted average azimuthal asymmetry coefficients $v_{n}$, for $n =2$, 4 and 6, with $0.3<\pt<3.0\GeVc$ and based on a second-order, $m=2$, reference frame are shown for three different methods as a function of centrality. The $v_2$ results are from Ref.~\cite{PhysRevC.87.014902} and included for completeness. (b) Results for the event plane and cumulant analyses for distributions based on higher-order, $m>2$, reference distributions.  Statistical (error bars) and systematic (light gray boxes) uncertainties are shown.}
    \label{fig:Fig8a}
  \end{center}
\end{figure*}

The centrality dependence of the yield-weighted average $v_{n}$ values with $0.3<\pt<3.0\GeVc$ is shown in Fig.~\ref{fig:Fig8} for the different methods. This is the \pt range for which a significant hydrodynamic-flow contribution is expected.  For completeness, the earlier $n=2$ results from Ref.~\cite{PhysRevC.87.014902} are also shown.  As noted for the \pt dependent results, a stronger centrality dependence is found
for analyses based on the $m=2$ reference plane. This is more clearly seen in Fig.~\ref{fig:Fig8a} where the analyses based on a second order, $m=2$ reference plane are shown in Fig.~\ref{fig:Fig8a}a and those based on higher order, $m>2$ reference planes are shown in Fig.~\ref{fig:Fig8a}b.   The $v_{6}\{\Psi_{6}\}$ results do not show a centrality dependence, although the large statistical uncertainties may mask a weak dependence.

\begin{figure*}[hbtp]
  \begin{center}
    \includegraphics[width=\cmsFigWidthStdWide]{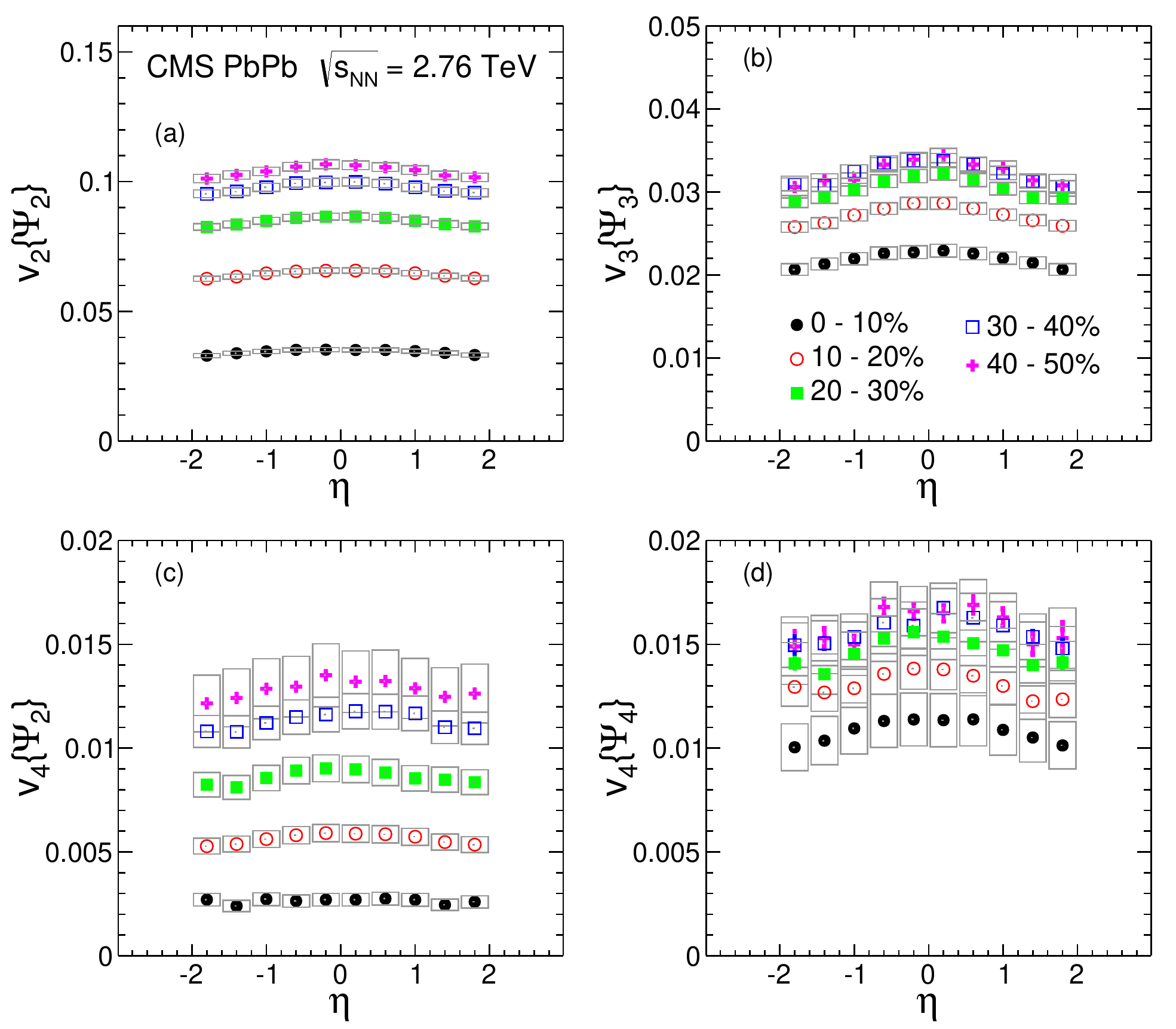}
    \caption{(Color online) Yield-weighted average azimuthal asymmetry coefficients $v_{n}$, for $n =2$--4,  with
    $0.3<\pt<3.0\GeVc$ are shown as a function of pseudorapidity $\eta$ for the indicated centrality ranges, as specified in percent. Statistical (error bars) and systematic (light gray boxes) uncertainties are shown. The $v_2$ results are from Ref.~\cite{PhysRevC.87.014902} and included for completeness.}
    \label{fig:Fig9}
  \end{center}
\end{figure*}

Figure~\ref{fig:Fig9} shows the pseudorapidity dependence for the yield-weighted average event plane $v_{n}$ values with $n=2$, 3, and 4, and with $0.3<\pt<3.0\GeVc$. The values
for the $n=5$ and 6 harmonics are found to be too small to establish their dependence on pseudorapidity.  The data are sorted into ten
pseudorapidity bins of $\Delta\eta=0.4$ spanning the range $-2.0<\eta<2.0$.  The distributions
have their maximum values near mid-rapidity, with the fractional decrease from $\abs{\eta}=0$ to $\abs{\eta}=2$ being similar for the different centrality ranges in a given harmonic.

\subsection{Comparison with other results}

\begin{figure*}[hbtp]
  \begin{center}
    \includegraphics[width=\cmsFigWidthStdWide]{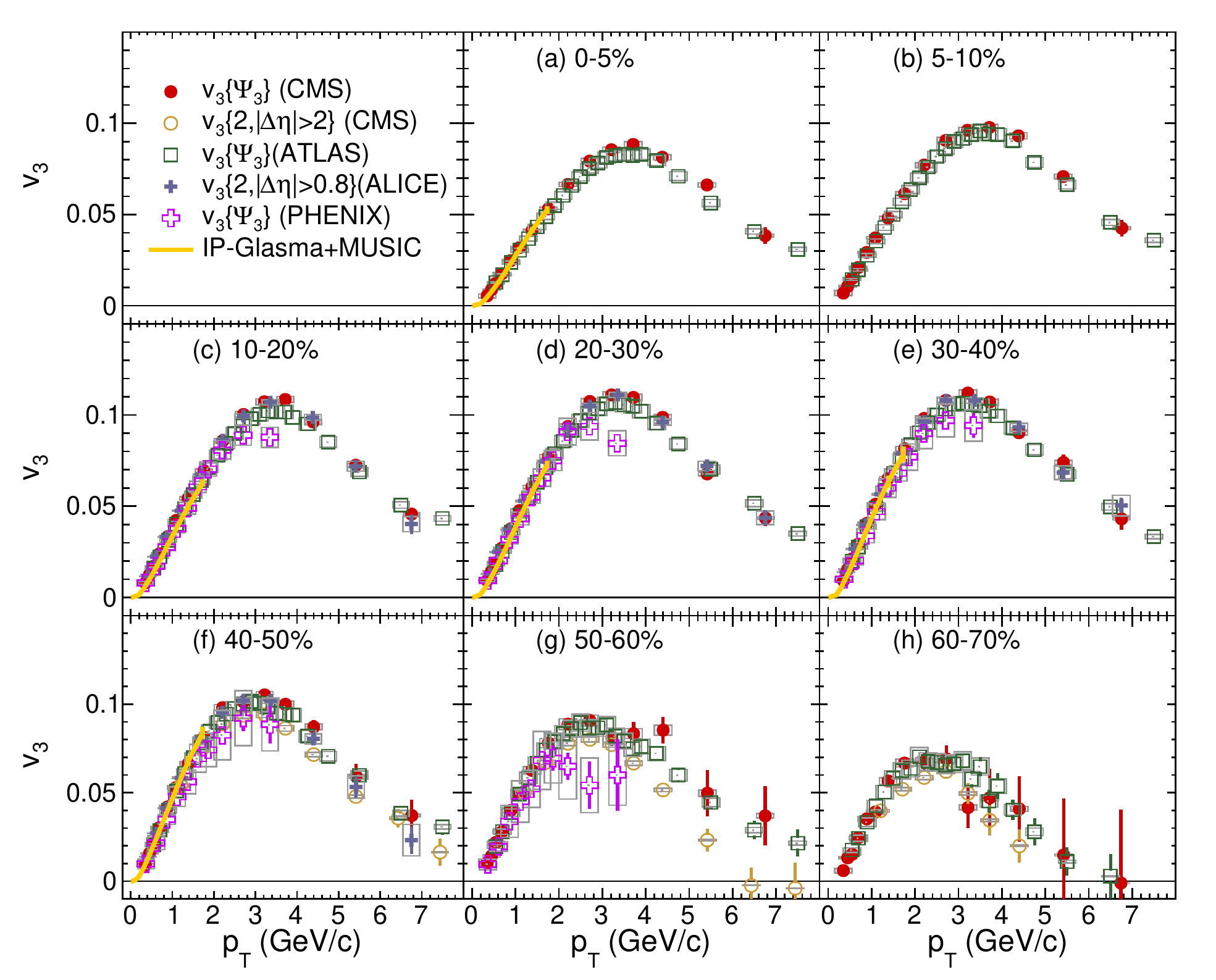}
    \caption{(Color online) Comparison of the $v_{3}$ results for PbPb collisions at $\sNN=2.76\TeV$ of the ALICE, ATLAS, and CMS Collaborations for the indicated centrality ranges, as specified in percent. The PHENIX results for AuAu collisions at $\sNN=200\GeV$ are also shown.  Statistical (error bars) and systematic (light gray boxes) uncertainties are indicated. References and experimental conditions are given in Table~\ref{tbl:expsum}. The predictions of the IP--Glasma+MUSIC model~\cite{PhysRevLett.110.012302} for PbPb collisions at $\sNN=2.76\TeV$  are shown by the solid lines in the 0--5\%, 10--20\%, 20--30\%, 30--40\%, and 40--50\% panels for $0<\pt<2\GeVc$.}
    \label{fig:Fig11}
  \end{center}
\end{figure*}

\begin{figure*}[hbtp]
  \begin{center}
    \includegraphics[width=\cmsFigWidthStdWide]{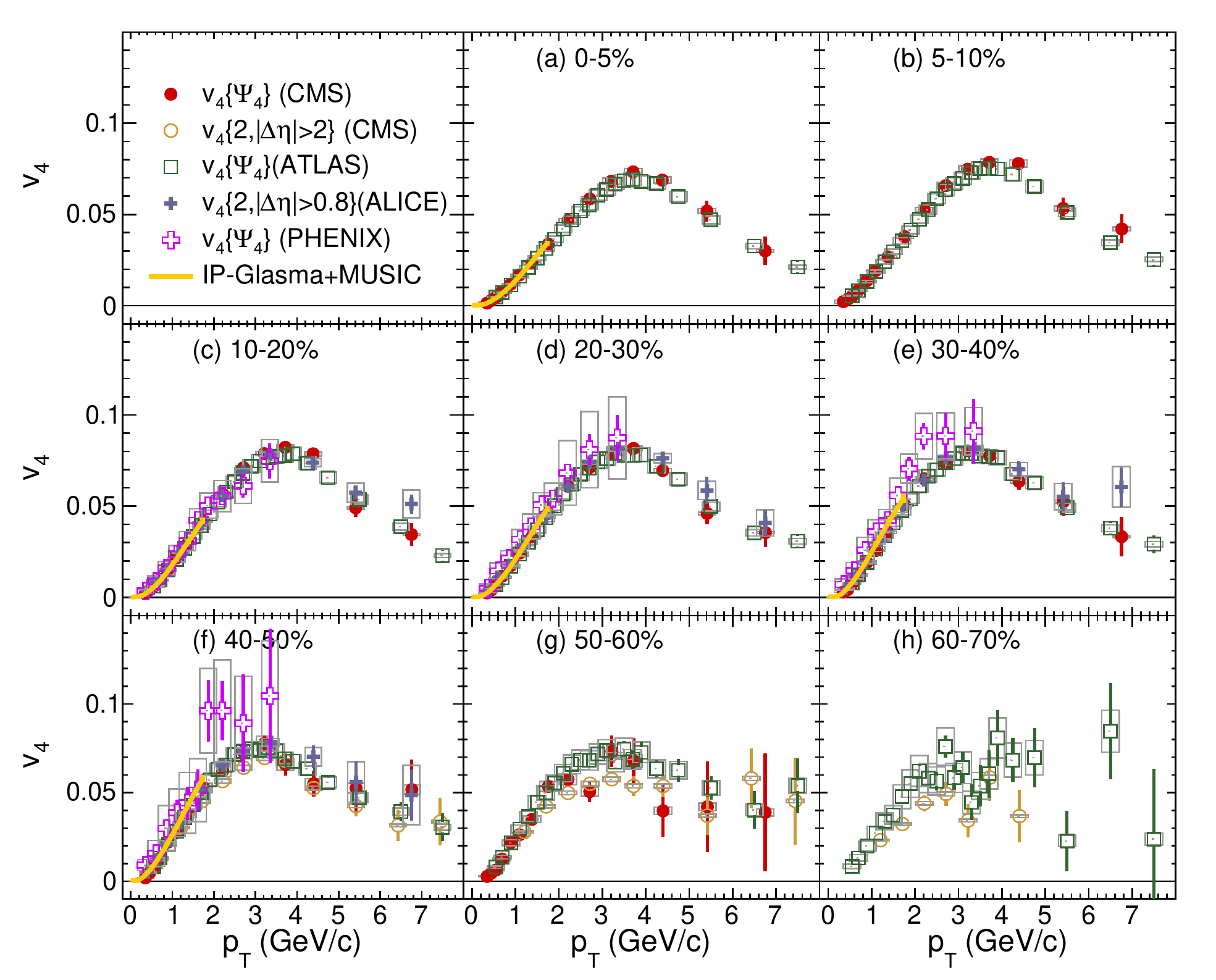}
    \caption{(Color online) Comparison of the $v_{4}$ results  for PbPb collisions at $\sNN=2.76\TeV$ of the ALICE, ATLAS, and CMS Collaborations for the indicated centrality ranges, as specified in percent. The PHENIX results for AuAu collisions at $\sNN=200\GeV$ are also shown. Statistical (error bars) and systematic (light gray boxes) uncertainties are indicated. References  and experimental conditions are given
    in Table~\ref{tbl:expsum}.  The predictions of the IP--Glasma+MUSIC model~\cite{PhysRevLett.110.012302}  for PbPb collisions at $\sNN=2.76\TeV$ are shown by the solid lines in the 0--5\%, 10--20\%, 20--30\%, 30--40\%, and 40--50\% panels for $0<\pt<2\GeVc$.}
    \label{fig:Fig12}
  \end{center}
\end{figure*}

\begin{figure*}[hbtp]
  \begin{center}
    \includegraphics[width=\cmsFigWidthStdWide]{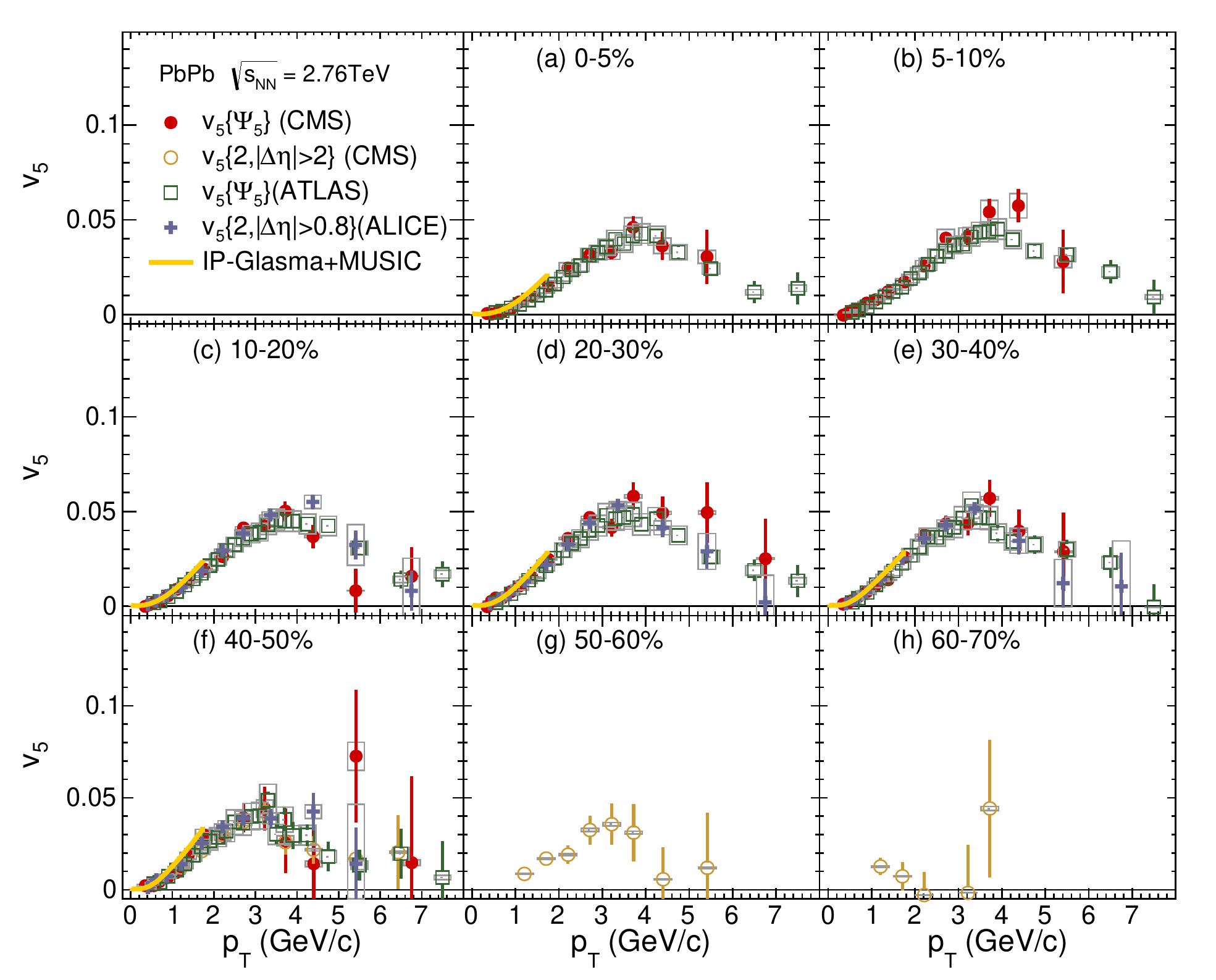}
    \caption{(Color online) Comparison of the $v_{5}$ results of the ALICE, ATLAS, and CMS Collaborations for the indicated centrality ranges, as specified in percent. Statistical (error bars) and systematic (light gray boxes) uncertainties are indicated. References and experimental conditions are given
    in Table~\ref{tbl:expsum}.  The predictions of the IP--Glasma+MUSIC model~\cite{PhysRevLett.110.012302} are shown by the solid lines in the 0--5\%, 10--20\%, 20--30\%, 30--40\%, and 40--50\% panels for $0<\pt<2\GeVc$.}
    \label{fig:Fig13}
  \end{center}
\end{figure*}

\begin{figure*}[hbtp]
  \begin{center}
    \includegraphics[width=\cmsFigWidthStdWide]{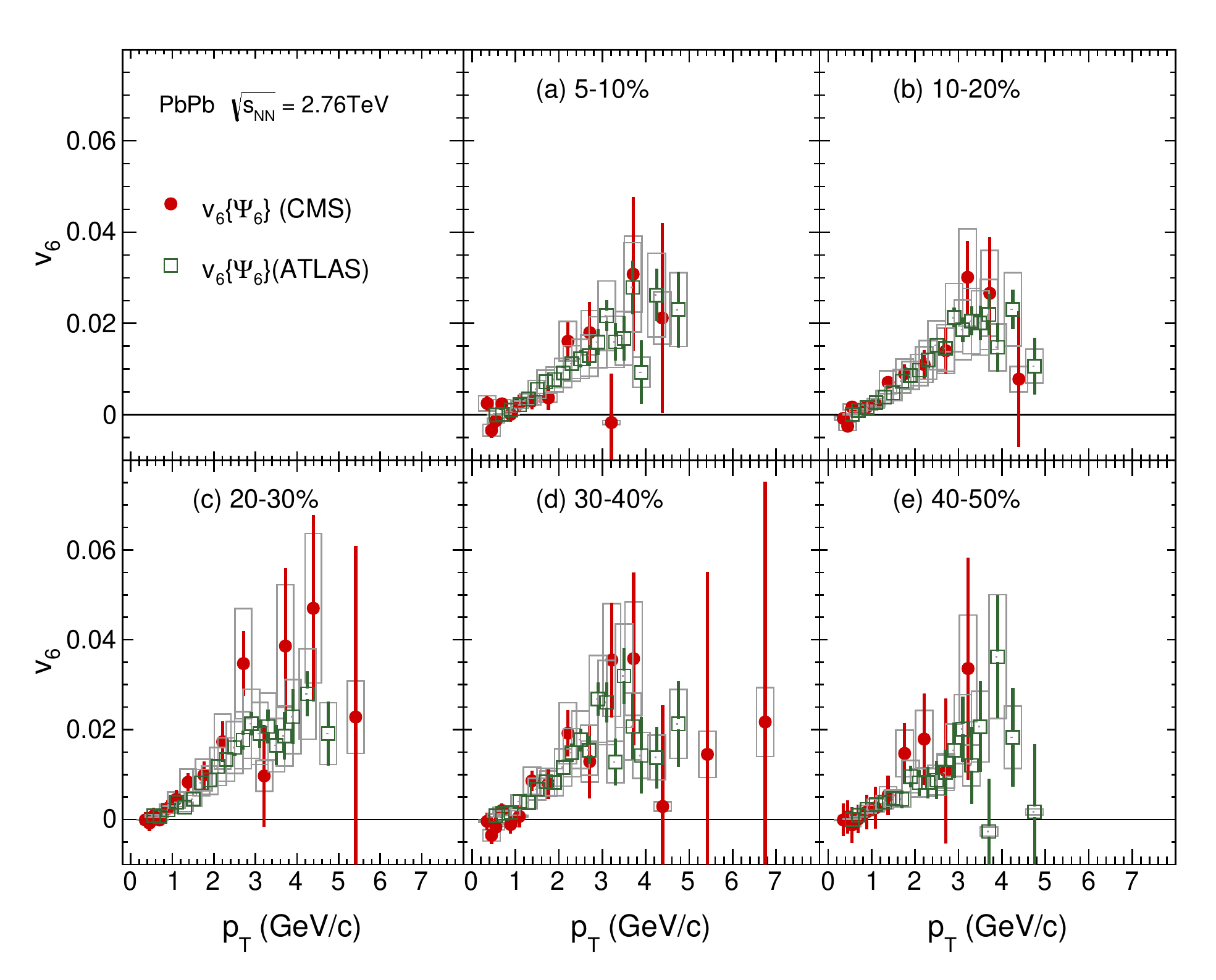}
    \caption{(Color online) Comparison of the $v_{6}$ results of the ATLAS and CMS Collaborations for the indicated centrality ranges, as specified in percent. Statistical (error bars) and systematic (light gray boxes) uncertainties are indicated. References  and experimental conditions are given
    in Table~\ref{tbl:expsum}. }
    \label{fig:Fig14}
  \end{center}
\end{figure*}

The current results extend and largely confirm previous results published by the ALICE~\cite{PhysRevLett.107.032301,PhysLettB.708.249} and ATLAS~\cite{PhysRevC.86.014907} Collaborations on higher-order harmonic correlations. Representative comparisons
of the CMS results with those of these other two collaborations are shown in Figs.~\ref{fig:Fig11} to \ref{fig:Fig14}. Corresponding results by the PHENIX Collaboration for AuAu collisions at $\sNN=200\GeV$ are also shown~\cite{PhysRevLett.107.252301}.  Differences
in the centrality and pseudorapidity ranges chosen by the different collaborations need to be considered
in comparing the results. Table~\ref{tbl:expsum} summarizes the experimental conditions for the different measurements.

\begin{table*}[hbtp]
\ifthenelse{\boolean{cms@external}}{}{\footnotesize}
\begin{center}
\topcaption{Summary of experimental conditions for the data shown in this report. The Figure(s)
column indicates the figures in this report where the data are shown. The \pt range for previously
published data corresponds to that shown in the original report.
}
\label{tbl:expsum}
\resizebox{\textwidth}{!}{
\begin{scotch}{ccccD{+}{\text{--}}{1,1}c}
\textbf{Method(s)}&Figure(s)& Collaboration &\multicolumn{1}{c}{$\eta$ \text{range}}&\multicolumn{1}{r}{\pt range (\GeVcns)}&\multicolumn{1}{c}{Reference}  \\
\hline
$v_{3}\{\Psi_{3}\}$,~$v_{4}\{\Psi_{4}\}$,~$v_{5}\{\Psi_{5}\}$, &\ref{fig:Fig4}--\ref{fig:Fig7}  &  CMS   &$\abs{\eta}<0.8$  &0.3+8.0 & this paper\\
$v_{4}\{\Psi_{2}\}$,~$v_{6}\{\Psi_{6}\}$,~$v_{6}\{\Psi_{2}\}$&&&&&\\
\noalign{\vskip 1.5ex}
$v_{3}\{4\}$  & \ref{fig:Fig4} &  CMS   &$\abs{\eta}<0.8$  &0.3+4.0 & this paper\\
$v_{4}\{5\}$ & \ref{fig:Fig5} &  CMS   &$\abs{\eta}<0.8$  &0.3+8.0 & this paper\\
$v_{4}\{\mathrm{LYZ}\}$, $v_{6}\{\mathrm{LYZ}\}$& \ref{fig:Fig5}, \ref{fig:Fig7} &  CMS   &$\abs{\eta}<0.8$  &0.3+8.0 & this paper\\
$v_{n}\{{2,\abs{\Delta\eta}>2}\}$ &\ref{fig:Fig4}--\ref{fig:Fig7}& CMS & $\abs{\eta}<2.5;~2<\abs{\Delta\eta}<4$ &1.0+20 & \cite{Eur.Phys.J.C72.2012}\\
$v_{n}\{{2,\abs{\Delta\eta}>2}\}$ &\ref{fig:Fig11}--\ref{fig:Fig13}& ALICE & $\abs{\eta}<1.0;~\abs{\Delta\eta}>0.8$ &0.25+15 & \cite{PhysLettB.708.249}\\
$v_{n}\{\Psi_{n}\}$&\ref{fig:Fig11}--\ref{fig:Fig14}& ATLAS & $\abs{\eta}<2.5$ &0.5+12 & \cite{PhysRevC.86.014907}\\
$v_{n}\{\Psi_{n}\}$&\ref{fig:Fig11}--\ref{fig:Fig12}& PHENIX & $\abs{\eta}<0.35$ &0.2+4.0 &  \cite{PhysRevLett.107.252301}\\
\end{scotch}
}
\end{center}
\end{table*}

Figure~\ref{fig:Fig11} compares results of CMS, ATLAS, and ALICE for the \pt dependent $v_{3}$ coefficient. The ATLAS results for
$v_{3}\{\Psi_{3}\}$ are consistently lower than the CMS results for all but the most peripheral centrality bin.  This is expected
based on the larger pseudorapidity range being used for the ATLAS measurement.  Good agreement is seen between the two particle
correlation results of the CMS and ALICE Collaborations.  This suggests that the pseudorapidity gap of $\abs{\Delta\eta}>0.8$ employed
by ALICE is already sufficient to remove most of the dijet contribution to these correlations.

The comparison of $v_{4}$  and $v_{5}$  values found by the three experiments lead to similarly consistent results. The CMS and ATLAS results are also very similar for $v_{6}\{\Psi_{6}\}(\pt)$ within the respective uncertainties, as seen in Fig.~\ref{fig:Fig14}. Figures~\ref{fig:Fig11}--\ref{fig:Fig13} also show the predictions of the IP--Glasma+MUSIC model of Ref.~\cite{PhysRevLett.110.012302}, as discussed in the next section. The model calculations cover the hydrodynamic-dominated region of the \pt distributions.

The lower energy $v_{3}\{\Psi_{3}\}$ and $v_{4}\{\Psi_{4}\}$ results of the PHENIX Collaboration for AuAu collisions at $\sNN=200\GeV$ are also shown in Figs.~\ref{fig:Fig11} and~\ref{fig:Fig12}, respectively.  The $n=3$ AuAu results are systematically lower than those obtained by the higher-energy LHC measurements, consistent with what was previously observed for the elliptic-flow, $n=2$ harmonic~\cite{PhysRevC.87.014902}.  A different picture is suggested by the $n=4$ distributions, where now the RHIC results are systematically higher than those observed at the LHC, although with large systematic uncertainties.

\section{Discussion}
\label{sec:Discussion}
\label{sec:discussion}
There is considerable interest in how the spatial anisotropies, as characterized by spatial anisotropy parameters $\epsilon_{n}$, created early in the collision of two ultra-relativistic heavy ions
get transformed into the experimentally observed azimuthal anisotropy of emitted particles~\cite{PhysRevC.81.054905,PhysRevC.81.014901,PhysRevC.81.061901,PhysRevC.82.064903,PhysRevC.84.034910,PhysRevC.84.064907,PhysRevC.83.044902,PhysRevC.84.024911,PhysRevC.85.014905,PhysRevLett.108.252301}.  The higher-order harmonics are expected to be more sensitive to the details of the collision geometry and its event-by-event fluctuations.
This section develops the scaling behavior of the experimental $v_{n}$ coefficients in terms of the Glauber model $\epsilon_{n}$ values and also explores the effect of fluctuations on the different analysis methods.

It is now recognized that the different experimental methods used in determining the $v_{n}$ coefficients are related differently to the underlying $\epsilon_{n}$ values.   For example, $v_{n}\{\Psi_{n}\}$ coefficients obtained with near-unity values for the event plane resolution factor $R$ are expected to scale with $\langle\epsilon_{n}\rangle$, whereas these coefficients scale with $\sqrt{\langle\epsilon_{n}^2\rangle}$ for lower values of $R$~\cite{PhysRevC.81.054905}. The two particle correlations are also expected to scale as $\sqrt{\langle\epsilon_{n}^2\rangle}$, whereas the $v_{n}\{4\}$ coefficient should scale as the fourth-order cumulant eccentricity, with~\cite{PhysRevC.84.024911}  $\epsilon_{2} \left\{ 4 \right\} = {\left( {{{\left\langle {\epsilon _2^2} \right\rangle }^2} - \left[ {\left\langle {\epsilon _2^4} \right\rangle  - {{\left\langle {\epsilon _2^2} \right\rangle }^2}} \right]} \right)^{1/4}}$.

\begin{figure*}[hbt]
  \begin{center}
    \includegraphics[width=\cmsFigWidthStdWide]{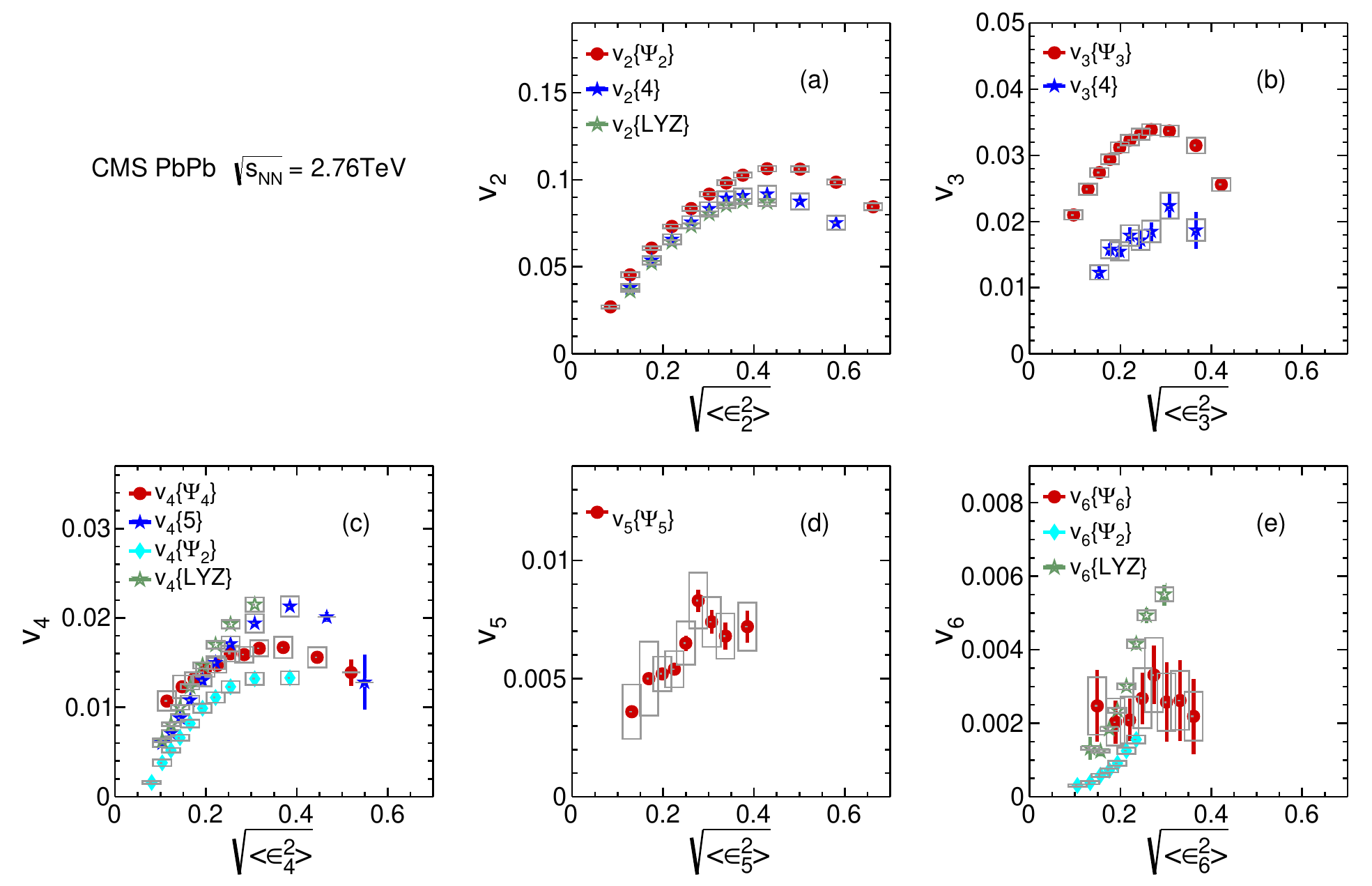}
    \caption{(Color online) Yield-weighted average azimuthal asymmetry parameters $v_{n}$, for $n = 2$--6 with $0.3<\pt<3.0\GeVc$, as a function of the corresponding Glauber model rms anisotropy parameters $\sqrt{<\epsilon_{n}^2>}$. The CMS $v_2$ results are from Ref.~\cite{PhysRevC.87.014902} and included for completeness.  The $v_{4}\{\Psi_{2}\}$ and $v_{6}\{\Psi_{2}\}$  results are plotted against $\sqrt{<\epsilon_{4,2}^{2}>}$ and $\sqrt{<\epsilon_{6,2}^{2}>}$, respectively,  as given in Table~\ref{ecc_rn}. Statistical (error bars) and systematic (light gray boxes) uncertainties are indicated.}
    \label{fig:Fig15}
  \end{center}
\end{figure*}

The details of the eccentricity scaling are model dependent and beyond the scope of this paper. However, to achieve an overview of the geometry scaled behavior, we present in Fig.~\ref{fig:Fig15} the yield-weighted average $v_{n}$ results of Fig.~\ref{fig:Fig8} as a function of the Glauber model
$\sqrt{\langle\epsilon_{n,m}^2\rangle}$ azimuthal asymmetries discussed in Section~\ref{sec:glauber}. In general, the $v_{n}$ coefficients are found to increase monotonically with the Glauber model eccentricities for the most central events, up to the maxima in the distributions shown in Fig.~\ref{fig:Fig8}, although large uncertainties are affecting the $n>4$ event plane results and some method differences are observed.  For $n=2$, both the event plane and four particle cumulant results show similar behavior for the most central events, with the overall magnitude of the $v_{2}\{4\}$ coefficients being smaller.   The observed difference is consistent with the fourth-order cumulant results scaling with the four particle cumulant eccentricity $\epsilon_2\{4\}$, as shown in Ref.~\cite{PhysRevC.87.014902}. A much larger difference is observed between the event plane and cumulant results for $n=3$, as would be expected if the odd harmonics are dominated by fluctuation effects, which are strongly suppressed in the multiparticle cumulant analysis.   The higher-order harmonic event plane results with $n=m$ show relatively weak scaling with the Glauber geometry, also suggesting significant fluctuation components.  For $n=4$ the harmonic components based on a second-order reference plane, as is the case for $v_{4}\{5\}$,  $v_{4}\{\Psi_{2}\}$, and $v_{4}\{\mathrm{LYZ}\}$, are found to have a much stronger dependence on
the Glauber eccentricity for more central events than is evident for the analysis based on the fourth-order event plane.

\begin{figure}[hbtp]
  \begin{center}
    \includegraphics[width=\cmsFigWidth]{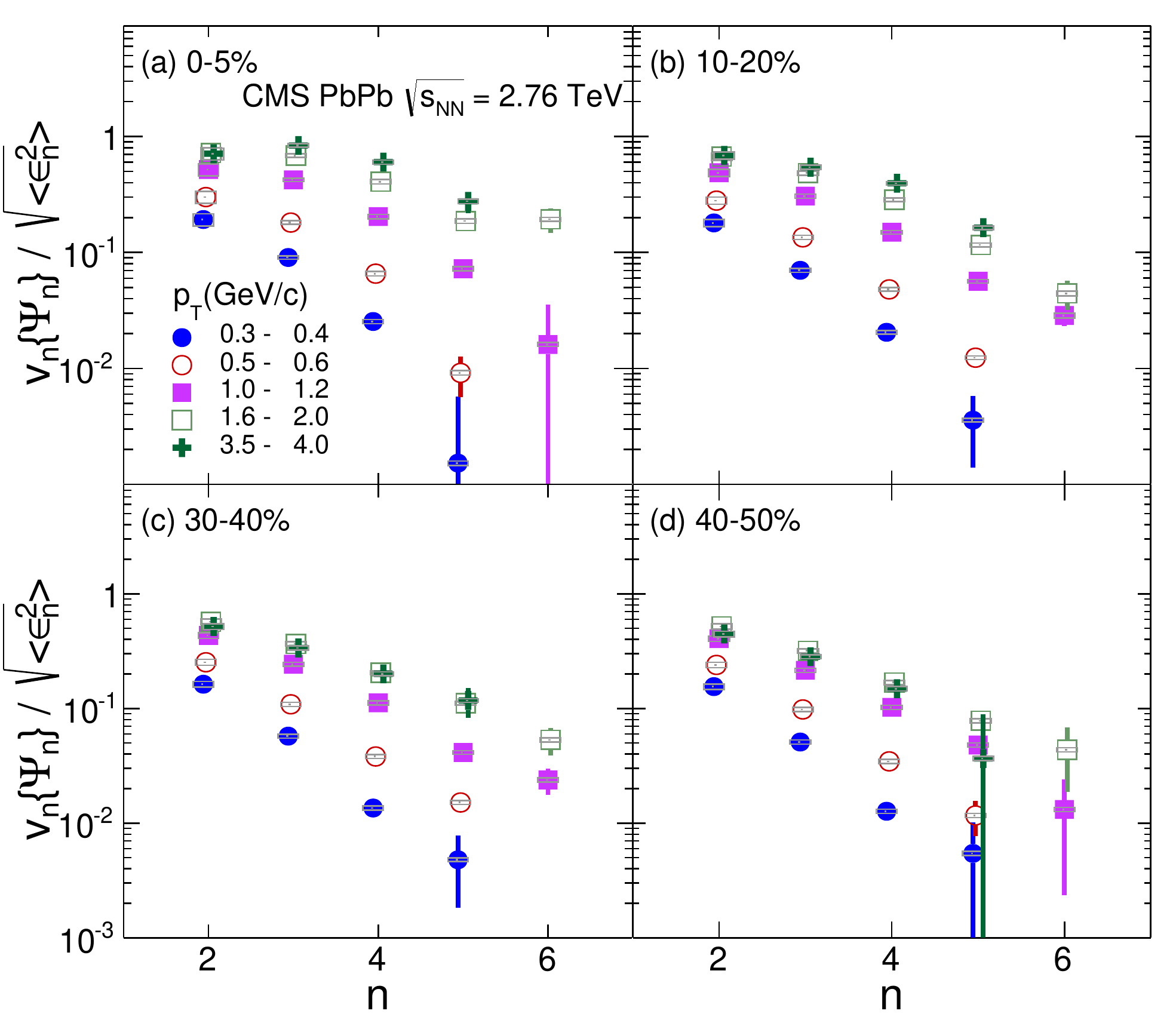}
    \caption{(Color online) The azimuthal asymmetry parameter $v_n$, scaled by the  corresponding Glauber model rms anisotropy parameter $\sqrt{\epsilon_n^2}$ for the indicated \pt and centrality ranges, as specified in percent, as a function of harmonic number $n$. Statistical (error bars) and systematic (light gray boxes) uncertainties are indicated. }
    \label{fig:Fig16}
  \end{center}
\end{figure}

Figure~\ref{fig:Fig16} shows the eccentricity scaled $v_{n}\{\Psi_{n}\}$ values as a function of the harmonic order $n$ for five different \pt ranges and
for four different centrality ranges. For all but the most central events, the $v_{n}$ values are found to decrease with increasing harmonic number.  The rate of this decrease is expected to be sensitive to the shear viscosity of the medium, which leads to greater damping of the higher-order harmonic anisotropies~\cite{PhysRevC.85.024901,PhysRevC.84.044912}. For the most central events, the scaled $v_{3}$ coefficient is found to become larger than $v_{2}$ for the highest \pt bin of $3.5<\pt<4.0\GeVc$.   Overall, the observed fall-off with harmonic order is very regular. The more central events demonstrate a fall-off that is steeper than an exponential in $n$.   For mid-central events, however, the fall-off appears to scale as an exponential in $n$.   Recent papers have suggested that the higher-order harmonic components may reflect a strong nonlinear response, particularly for noncentral events, with the higher-order harmonics dependent on mixtures of lower-order eccentricities~\cite{PhysRevC.85.024908,PhysRevC.86.044908}.

\begin{figure}[hbtp]
  \begin{center}
    \includegraphics[width=\cmsFigWidthStd]{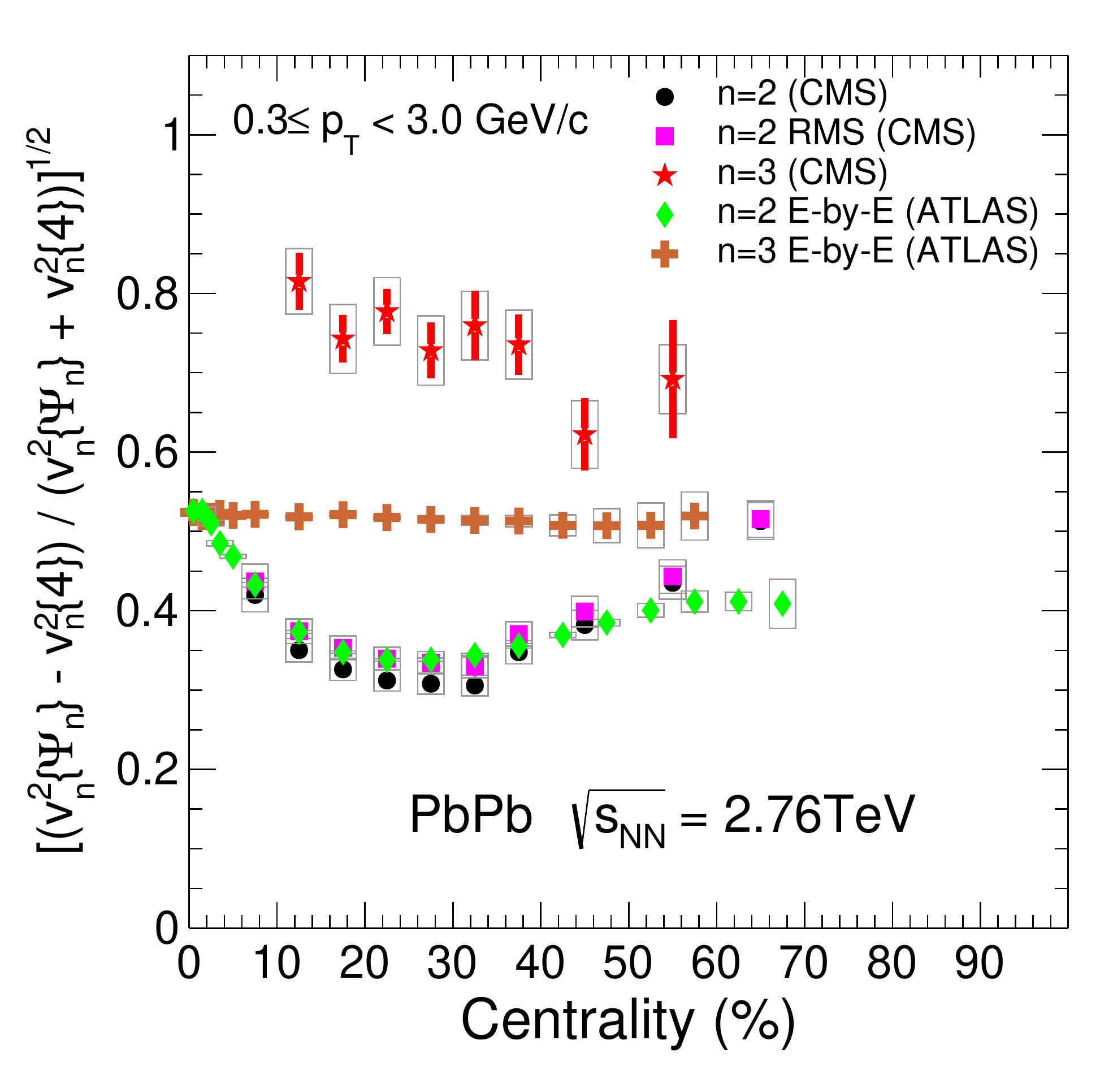}
    \caption{(Color online) Estimate of the event-by-event $v_{n}$ fluctuations in $v_n$ for $n=$ 2 (filled circles) and 3 (filled stars) as discussed in the text.  The results where the event plane $v_2$ values are adjusted to their corresponding rms values, as discussed in the text, are indicated by the filled boxes and labelled $n = 2\,\mathrm{RMS}$.   The results of event-by-event-fluctuation (E-by-E) analyses by the  ATLAS Collaboration for the $n$ = 2 and 3 harmonics~\cite{arXiv:1305.2942} are also shown. Statistical (error bars) and systematic (light gray boxes) uncertainties are indicated. The systematic uncertainties for the ATLAS results  correspond to the mid-point of the uncertainty range indicated by the ATLAS Collaboration. }.
    \label{fig:Fig17}
  \end{center}
\end{figure}

Event-by-event fluctuations in the location of the participant nucleons can have a large and
method dependent influence on the harmonic coefficients~\cite{PhysRevC.77.014906,PhysRevC.80.014904}.
Expressing the fluctuations in terms of the azimuthal anisotropy in the participant plane $v$, where the harmonic number
is suppressed, the magnitude of the fluctuations is given by
$\sigma _v^2 \equiv \left\langle {{v^2}} \right\rangle  - {\left\langle v \right\rangle ^2}$.
It can then be shown~\cite{PhysRevC.80.014904} that to leading order in
$\sigma_v$,
two and four particle correlations are affected differently, with
\begin{equation}
\label{eqn:Eqn11}
v{\left\{ 2 \right\}^2} = \left\langle {{v^2}} \right\rangle  = {\left\langle v \right\rangle ^2} + \sigma _v^2
\end{equation}
and
\begin{equation}
\label{eqn:Eqn12}
v{\left\{ 4 \right\}^2} = {\left( {2{{\left\langle {{v^2}} \right\rangle }^2} - \left\langle {{v^4}} \right\rangle } \right)^{1/2}} \approx {\left\langle v \right\rangle ^2} - \sigma _v^2.
\end{equation}
The event plane method leads to an intermediate value, with
\begin{equation}
\label{eqn:Eqn13}
v{\left\{ {{\mathrm{EP}}} \right\}^2} = {\left\langle v \right\rangle ^2} + \left( {\alpha  - 1} \right)\sigma _v^2 ,
\end{equation}
where $\alpha$ is a parameter between 1 and 2 that is determined empirically in terms of the event plane resolution factor $R$
(Fig.~\ref{fig:Fig2})~\cite{PhysRevC.77.014906}. Multiparticle correlations with greater than four particles are
expected to give results similar to those of four particle correlations. For harmonics with $n>2$, the event plane resolutions lead to
$v\left\{ {{\mathrm{EP}}} \right\} \approx v\left\{ 2 \right\}$ using the parameterization of $\alpha$
in terms of the event plane resolution $R$ given in Ref.~\cite{PhysRevC.77.014906}.
Based on the larger $R$ values observed for the CMS mid-central events with $n=2$,  the ratio of $v_2\{2\}$ to $v_2\{\Psi_2\}$ is expected to be about 1.02~\cite{PhysRevC.80.014904}. Motivated by Eqs.~(\ref{eqn:Eqn11}) and~(\ref{eqn:Eqn12}),
as well as the approximate equality of $v_n\{2\}$ and $v_n\{\Psi_n\}$, Fig.~\ref{fig:Fig17} shows the ratio
\begin{equation}
\label{eqn:Eqn14}
{\left[ {\left( {v_n^2\left\{ {{\Psi _n}} \right\} - v_n^2\left\{ {{\text{Cum}}} \right\}} \right)
/\left( {v_n^2\left\{ {{\Psi _n}} \right\} + v_n^2\left\{ {{\text{Cum}}} \right\}} \right)} \right]^{1/2}}
\end{equation}
for $n=2$ and 3.
The cumulant multiparticle correlation harmonics are indicated by $v_n\{\text{Cum}\}$,
using four particle correlations for both $n=2$ and 3.
If the magnitude of the fluctuations is relatively small compared to the corresponding harmonic anisotropy term,
this ratio should approach ${\sigma _v}/\left\langle v \right\rangle $.
The points shown as solid squares in Fig.~\ref{fig:Fig17} are obtained by scaling the observed $n=2$ event plane results to the limiting value associated with poor event plane resolution (i.e., $\sqrt{\langle v_n^2 \rangle }$).
The resulting elliptic-flow ($n=2$) fluctuation fraction near 0.4 is similar to that observed at RHIC for AuAu collisions at
$\sNN  = 200\GeV$
~\cite{PhysRevLet.14.142301,PhysRevC.86.014904}, although the more recent STAR results~\cite{PhysRevC.86.014904} are systematically higher than those observed at the LHC.
The relative fluctuations are much larger for the $n=3$ harmonic than for the elliptic flow, $n=2$ harmonic.

In a recent event-by-event fluctuation analysis by the ATLAS Collaboration~\cite{arXiv:1305.2942}, the  ${\sigma _v}/\left\langle v \right\rangle $ ratio
was measured directly by unfolding the event-by-event $v_n$ distributions with the multiplicity dependence of the measurements.  These results are shown in Fig.~\ref{fig:Fig17} for the $n=2$ and 3 harmonics.  The CMS and ATLAS $n=2$ results are in good agreement except for the most peripheral bins, where the CMS results are higher.  The ATLAS analysis for the $n=3$ harmonic leads to a relatively constant value of ${\sigma _v}/\left\langle v \right\rangle $  with $N_\text{part}$ of approximately 0.53.   For this higher harmonic the leading-order assumption made for Eqs.~(\ref{eqn:Eqn11})
and~(\ref{eqn:Eqn12}) is violated.

\begin{figure}[hbtp]
  \begin{center}
    \includegraphics[width=\cmsFigWidthStd]{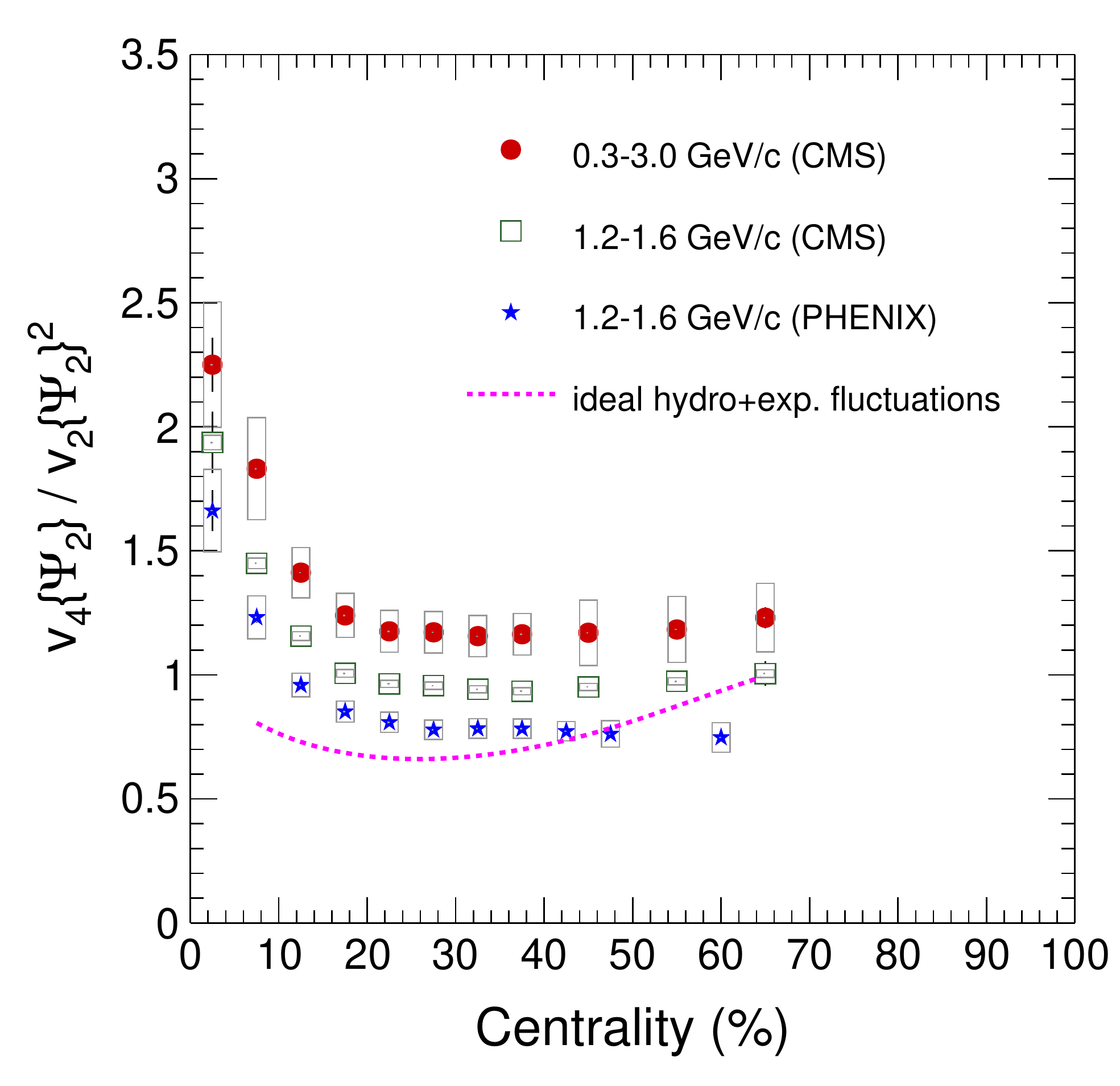}
    \caption{(Color online) The yield-weighted average values of the ratio $v_{4}/v_{2}^{2}$ as a function of centrality
    for  $0.3<\pt<3.0\GeVc$ (filled circles) and  $1.2<\pt < 1.6\GeVc$ (open squares) are shown for PbPb collisions at $\sNN=2.76\TeV$. The PHENIX results (stars) for $1.2<\pt < 1.6\GeVc$ are also shown for AuAu collisions at $\sNN=200$\GeV~\cite{PhysRevLett.105.062301}. The dotted line is based on the method presented in Ref.~\cite{PhysRevC.81.014901} that allows the expected $v_{4}/v_{2}^{2}$ ratio for an event plane analysis to be
calculated based on the ideal hydrodynamics limit of 0.5 and the observed relative fluctuation ratio $\sigma/\langle v_n \rangle$, as discussed in the text. Statistical (error bars) and systematic (light gray boxes) uncertainties are indicated. }
    \label{fig:Fig18}
  \end{center}
\end{figure}

Another ratio that has received considerable attention in characterizing the
azimuthal asymmetry is $v_{4}/v_{2}^{2}$~\cite{PhysRevC.68.031902,PhysRevC.69.051901,PhysRecC.75.054906,PhysRevLett.105.062301,PhysRevC.81.061901,PhysLettB.627.49,PhysRevC.81.014901,PhysRevC.85.024901}, where the $n$=2 and 4 harmonics are determined with respect to the elliptic-flow event plane.   It is now recognized that this ratio, with a value  of 0.5 obtained through ideal hydrodynamics~\cite{PhysRevC.69.051901}, is strongly affected by flow fluctuations and nonflow correlations. The comparisons of  theory to the experimental results need to account for how the results of the different analysis methods relate to the  event-by-event $v_{n}$ asymmetry~\cite{PhysRevC.81.014901}.   Figure~\ref{fig:Fig18} shows the ratio
$v_{4}\{\Psi_{2}\}/ v_{2}^{2}\{\Psi_{2}\}$ for two different \pt ranges as a function of centrality.   In both cases the ratio
initially decreases, but then remains relatively constant for centralities greater than $\approx 20\%$. The ratio using the yield-weighted average
$v_{n}$ values over the larger \pt range of $0.3<\pt < 3.0\GeVc$ is systematically larger than that found
in the \pt range of $1.2<\pt < 1.6\GeVc$.  The AuAu results obtained at $\sNN=~200\GeV$ by the PHENIX Collaboration for the range $1.2<\pt < 1.6\GeVc$ are also shown.  The CMS results are systematically higher by about 10\%.

It has been shown that if the harmonic coefficients reflect ideal hydrodynamics with additional participant fluctuation effects, then the expected ratio is given by~\cite{PhysRevC.81.014901}
\begin{equation}
\label{eqn:Eqn15}
\frac{{{v_4}\left\{ {{\Psi _2}} \right\}}}{{{v_2}{{\left\{ {{\Psi _2}} \right\}}^2}}} = \frac{1}{2}\left[ {1 + \beta {{\left( {\frac{{{\sigma _v}}}{{\left\langle v \right\rangle }}} \right)}^2}} \right] ,
\end{equation}
where $\beta$ depends on the event plane resolution parameter $R$. The dashed line in Fig.~\ref{fig:Fig18} shows this result using a smooth fit to the ${\sigma _v}/\left\langle v \right\rangle$ behavior found in Fig.~\ref{fig:Fig17}  and the subevent results shown in Fig. 6 of Ref.~\cite{PhysRevC.81.014901} for $\beta$.  The ideal hydrodynamic picture underestimates the observed $v_{4}\{\Psi_{2}\}/ v_{2}^{2}\{\Psi_{2}\}$ ratio at the LHC. Similar values might be expected for this ratio at
RHIC and LHC energies based on the similar values deduced for
${\sigma _v}/\left\langle v \right\rangle$ in experiments at the two facilities.

\begin{figure}[hbtp]
  \begin{center}
    \includegraphics[width=\cmsFigWidthStd]{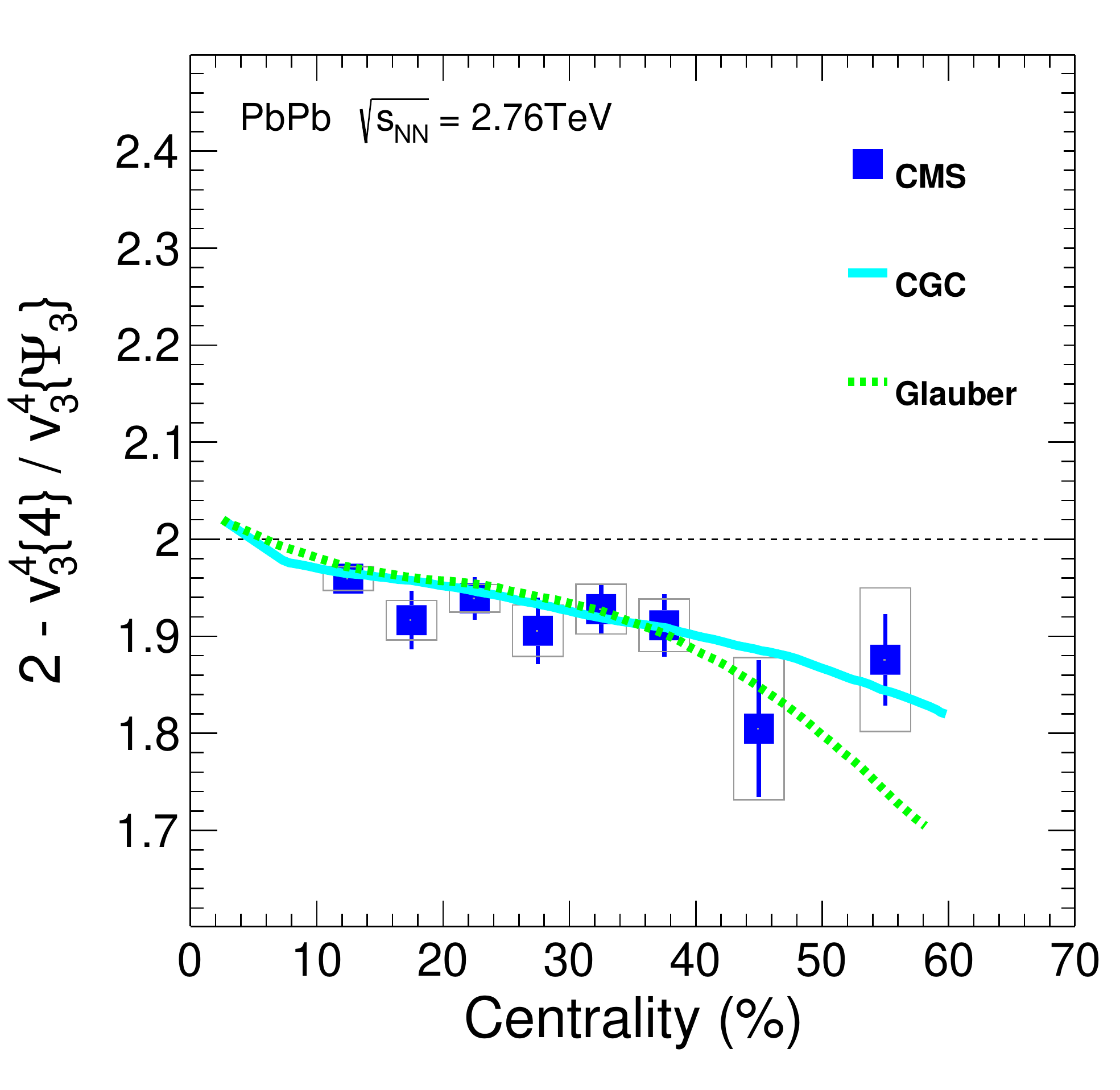}
    \caption{(Color online) Measurements of $2 - v_{3}^{4}\{4\}/v_{3}^{4}\{\Psi_{3}\}$ versus centrality.
    The CGC and Glauber model calculations are from Ref.~\cite{PhysRevC.84.034910}. Statistical (error bars) and systematic (light gray boxes) uncertainties are indicated. }
    \label{fig:Fig19}
  \end{center}
\end{figure}

In general, the flow harmonics are related to the initial state eccentricities through proportionality constants that depend on medium properties. It has been suggested that a greater sensitivity to initial state conditions might be achieved by studying the ratios of azimuthal anisotropy coefficients based on correlations of different numbers of particles or of mixed order~\cite{PhysRevC.84.034910,arXiv:1106.4940}. One such ratio based on correlations with different numbers of particles is given by $\left( {2v_n^4\left\{ 2 \right\} - v_n^4\left\{ 4 \right\}} \right)/v_n^4\left\{ 2 \right\}$~\cite{arXiv:1106.4940}.  Figure~\ref{fig:Fig19}  shows the quantity $2 - v_{3}^{4}\{4\}/v_{3}^{4}\{\Psi_{3}\}$ as a function of centrality, together with the corresponding CGC and Glauber model predictions~\cite{arXiv:1106.4940} . The yield-weighted average flow coefficients with $0.3<\pt<3.0\GeVc$ are used in determining the results from this analysis. The event plane results for  $v_{3}\{\Psi_{3}\}$ correspond to relatively low values of the resolution correction factor (see Fig.~\ref{fig:Fig2}) and, consequently, are expected to give similar results to $v_{3}\{2\}$.   Within the current experimental uncertainties, it is not possible to clearly state which model works best.  However, the results do suggest that a better determination of this ratio could help establish whether the initial state geometry is better described in a Glauber or CGC picture.

The STAR Collaboration at RHIC has recently released results for third-harmonic flow of charged particles from AuAu collisions at $\sNN=200\GeV$~\cite{arXiv:1301.2187} and has presented them in terms of the ratio plotted in Fig.~\ref{fig:Fig19}.
Their results are consistent with having a constant value near 2 for the full centrality range, although with relatively large uncertainties.  The four particle cumulant suppresses Gaussian fluctuations, and it is suggested that the larger multiplicities achieved at LHC energies may make these higher-energy results more sensitive to such fluctuations~\cite{arXiv:1301.2187}.

In a recent Letter~\cite{PhysRevLett.110.012302} the higher-order harmonic coefficients have been predicted based on a calculation that uses the impact parameter dependent Glasma (IP--Glasma) model~\cite{PhysRevLett.108.252301,PhysRevC.86.034908} to determine the early time evolution, and then switches to a relativistic hydrodynamic description using a (3+1)-dimensional relativistic viscous hydrodynamic simulation (MUSIC)~\cite{PhysRevLett.106.042301}. The IP--Glasma model includes not only the quantum fluctuations associated with the distribution of nucleons, as reflected in Glauber model calculations, but also fluctuations in the color charge distributions inside a nucleon.  These color charge fluctuations result in smaller-scale structure in the initial energy density profile than would be present if only sources of nucleon dimensions are considered.  The results are shown by the curves in Figs.~\ref{fig:Fig11}--\ref{fig:Fig13}.  Good agreement is found with the observed $v_{n}\{\pt\}$ behavior in the lower-\pt ranges.

\section{Summary}
\label{sec:Conclusions}
\label{sec:conclusions}
Results from the CMS Collaboration have been presented on higher-order harmonic anisotropies of charged particles for PbPb collisions at
$\sNN=2.76\TeV$. The harmonic coefficients $v_{n}$ have been studied as a function of transverse momentum ($0.3<\pt < 8.0\GeVc$), centrality (0--70\%), and pseudorapidity ($\abs{\eta} < 2.0$) using the event plane, cumulant, and Lee--Yang zeros methods.  The event plane method results are obtained with a pseudorapidity gap of at least three
units, with the event plane determined in the range $3<\abs{\eta}<5$,   suppressing the contribution of nonflow effects.

Comparisons of the event plane results with those of the cumulant and Lee--Yang zeros analyses suggest a strong influence of initial state fluctuations on the azimuthal anisotropies.  The weak centrality dependence
found for the event plane results based on event planes of harmonic order greater than two is
also consistent with the presence of a strong fluctuation component.
The pseudorapidity dependence of the higher-order azimuthal anisotropy parameters based on the event plane method is similar to that observed for elliptic flow,
with only a modest decrease from the mid-rapidity values out to the limits of the measurement at $\abs{\eta}=2.0$.   The mid-rapidity values are compared to those obtained by the ALICE~\cite{PhysLettB.708.249,PhysRevLett.107.032301} and ATLAS~\cite{PhysRevC.86.014907} Collaborations, and found to be in good agreement.  A comparison is also done with lower-energy AuAu measurements by the PHENIX Collaboration at $\sNN=200\GeV$, with only small differences found with the much higher energy LHC data.

The results obtained for $v_3$ are compared to predictions from
both the  CGC and Glauber models. Both of these initial state models are found to be consistent with
the data.  It is noted that a calculation that employs IP--Glasma-model initial conditions for the early time evolution, followed by a viscous hydrodynamic development of the plasma, is quite successful in reproducing the observed $v_{n}(\pt)$ results in the low-\pt, flow-dominated region~\cite{PhysRevLett.110.012302}.

The measurements presented in this paper help to further establish the pattern of azimuthal particle emission at LHC energies.  Recent theoretical investigations have significantly increased our understanding of the initial conditions and hydrodynamics that lead to the experimentally observed asymmetry patterns.  However, further calculations are needed to fully explain the method dependent differences seen in the data for the anisotropy harmonics. These differences can be attributed to the role of fluctuations in the participant geometry. Understanding the role of these fluctuations is necessary in order to establish the initial state of the created medium, thereby allowing for an improved determination of its properties. The current results are directly applicable to the study of the initial spatial anisotropy, time development, and shear viscosity of the medium formed in ultra-relativistic heavy-ion collisions.
\section*{Acknowledgments}
\hyphenation{Bundes-ministerium Forschungs-gemeinschaft Forschungs-zentren} We congratulate our colleagues in the CERN accelerator departments for the excellent performance of the LHC and thank the technical and administrative staffs at CERN and at other CMS institutes for their contributions to the success of the CMS effort. In addition, we gratefully acknowledge the computing centres and personnel of the Worldwide LHC Computing Grid for delivering so effectively the computing infrastructure essential to our analyses. Finally, we acknowledge the enduring support for the construction and operation of the LHC and the CMS detector provided by the following funding agencies: the Austrian Federal Ministry of Science and Research; the Belgian Fonds de la Recherche Scientifique, and Fonds voor Wetenschappelijk Onderzoek; the Brazilian Funding Agencies (CNPq, CAPES, FAPERJ, and FAPESP); the Bulgarian Ministry of Education, Youth and Science; CERN; the Chinese Academy of Sciences, Ministry of Science and Technology, and National Natural Science Foundation of China; the Colombian Funding Agency (COLCIENCIAS); the Croatian Ministry of Science, Education and Sport; the Research Promotion Foundation, Cyprus; the Ministry of Education and Research, Recurrent financing contract SF0690030s09 and European Regional Development Fund, Estonia; the Academy of Finland, Finnish Ministry of Education and Culture, and Helsinki Institute of Physics; the Institut National de Physique Nucl\'eaire et de Physique des Particules~/~CNRS, and Commissariat \`a l'\'Energie Atomique et aux \'Energies Alternatives~/~CEA, France; the Bundesministerium f\"ur Bildung und Forschung, Deutsche Forschungsgemeinschaft, and Helmholtz-Gemeinschaft Deutscher Forschungszentren, Germany; the General Secretariat for Research and Technology, Greece; the National Scientific Research Foundation, and National Office for Research and Technology, Hungary; the Department of Atomic Energy and the Department of Science and Technology, India; the Institute for Studies in Theoretical Physics and Mathematics, Iran; the Science Foundation, Ireland; the Istituto Nazionale di Fisica Nucleare, Italy; the Korean Ministry of Education, Science and Technology and the World Class University program of NRF, Korea; the Lithuanian Academy of Sciences; the Mexican Funding Agencies (CINVESTAV, CONACYT, SEP, and UASLP-FAI); the Ministry of Science and Innovation, New Zealand; the Pakistan Atomic Energy Commission; the Ministry of Science and Higher Education and the National Science Centre, Poland; the Funda\c{c}\~ao para a Ci\^encia e a Tecnologia, Portugal; JINR (Armenia, Belarus, Georgia, Ukraine, Uzbekistan); the Ministry of Education and Science of the Russian Federation, the Federal Agency of Atomic Energy of the Russian Federation, Russian Academy of Sciences, and the Russian Foundation for Basic Research; the Ministry of Science and Technological Development of Serbia; the Secretar\'{\i}a de Estado de Investigaci\'on, Desarrollo e Innovaci\'on and Programa Consolider-Ingenio 2010, Spain; the Swiss Funding Agencies (ETH Board, ETH Zurich, PSI, SNF, UniZH, Canton Zurich, and SER); the National Science Council, Taipei; the Thailand Center of Excellence in Physics, the Institute for the Promotion of Teaching Science and Technology and National Electronics and Computer Technology Center; the Scientific and Technical Research Council of Turkey, and Turkish Atomic Energy Authority; the Science and Technology Facilities Council, UK; the US Department of Energy, and the US National Science Foundation.

Individuals have received support from the Marie-Curie programme and the European Research Council (European Union); the Leventis Foundation; the A. P. Sloan Foundation; the Alexander von Humboldt Foundation; the Belgian Federal Science Policy Office; the Fonds pour la Formation \`a la Recherche dans l'Industrie et dans l'Agriculture (FRIA-Belgium); the Agentschap voor Innovatie door Wetenschap en Technologie (IWT-Belgium); the Ministry of Education, Youth and Sports (MEYS) of Czech Republic; the Council of Science and Industrial Research, India; the Compagnia di San Paolo (Torino); and the HOMING PLUS programme of Foundation for Polish Science, cofinanced from European Union, Regional Development Fund.

\bibliography{auto_generated}   
\cleardoublepage
\ifthenelse{\boolean{cms@external}}{
\appendix*
\section{Systematic uncertainties}
}
{
\appendix
\section{Systematic uncertainties}
}
\label{sec:Appendix}
The systematic uncertainties for the results presented in this paper are given in Tables~\ref{tab:sys_v3_EP_combined} to \ref{tab:sys_v6_LYZ_combined}.

\begin{table}[h!tbp]
\begin{center}
\topcaption{Systematic uncertainties in the $v_{3}\{\Psi_{3}\}$ values as a function of centrality in percent.
Common uncertainties are  shown at the top of the table, followed by those specific to the differential (\pt dependent) and
integral ($\abs{\eta}$ dependent) measurements.}
\label{tab:sys_v3_EP_combined}
\begin{scotch}{p{3cm}cccc}
\textbf{Source} &&\multicolumn{3}{c}{\textbf{Centrality}}\\
& &0--10\% &10--50\% &50--70\% \\
\hline
Particle &  & 0.5 & 0.5 & 0.5 \\
Composition \\
Centrality &  & 1.0 & 1.0 & 1.0 \\
Determination  \\
Resolution &  & 1.0 & 1.0 & 3.0 \\
Correction  \\
\hline \textbf{[Differential]}&\pt (\GeVcns)\\ \hline
Track Quality & 0.3--0.4 &  20 &  10 &  20 \\
Requirements  & 0.4--0.8 & 3.0 & 2.0 & 2.0 \\
& 0.8--8.0 & 1.0 & 1.0 & 1.0 \\
\hline
\textbf{Total ${(\pt)}$ } & 0.3--0.4 &  20 &  10 &  20 \\
& 0.4--0.8 & 3.4 & 2.5 & 3.8 \\
& 0.8--8.0 & 1.8 & 1.8 & 3.4 \\
\hline \textbf{[Integral]}&${\abs{\eta}}$&\\ \hline
Track Quality & 0.0--1.6 & 3.0 & 2.0 & 2.0 \\
Requirements  & 1.6--2.4 & 6.0 & 4.0 & 4.0 \\
\hline
\textbf{Total ${(\abs{\eta})}$ } & 0.0--1.6 & 3.4 & 2.5 & 3.8 \\
& 1.6--2.4 & 6.2 & 4.3 & 5.1 \\
\end{scotch}
\end{center}
\end{table}

\begin{table}[tbp]
\begin{center}
\topcaption{Systematic uncertainties in the $v_{4}\{\Psi_{4}\}$ values as a function of centrality in percent.
Common uncertainties are  shown at the top of the table, followed by those specific to the differential ($\pt$ dependent) and
integral ($\abs{\eta}$ dependent) measurements.}
\label{tab:sys_v4_EP_combined}
\begin{scotch}{p{3cm}cccc}
\textbf{Source} &&\multicolumn{3}{c}{\textbf{Centrality}}\\
& &0--10\% &10--40\% &40--60\% \\
\hline
Particle &  & 0.5 & 0.5 & 0.5 \\
Composition \\
Centrality &  & 1.0 & 1.0 & 1.0 \\
Determination  \\
Resolution &  & 2.0 & 2.0 & 5.0 \\
Correction  \\
\hline\textbf{[Differential]}&\pt (\GeVcns)&\\ \hline
Track Quality & 0.3--0.4 &  40 &  10 &  10 \\
Requirements  & 0.4--0.8 & 6.0 & 4.0 & 4.0 \\
& 0.8--8.0 & 1.0 & 1.0 & 1.0 \\
\hline\textbf{Total} ${(\pt)}$ & 0.3--0.4 &  40 &  10 &  11 \\
& 0.4--0.8 & 6.4 & 4.6 & 6.5 \\
& 0.8--8.0 & 2.5 & 2.5 & 5.2 \\
\hline \textbf{[Integral]}&${\abs{\eta}}$&\\ \hline
Track Quality & 0.0--1.6 &  10 & 5.0 & 5.0 \\
Requirements  & 1.6--2.4 &  20 & 8.0 & 8.0 \\
\hline \textbf{Total} ${(\abs{\eta})}$ & 0.0--1.6 &  10 & 5.5 & 7.2 \\
& 1.6--2.4 &  20 & 8.3 & 9.5 \\
\end{scotch}
\end{center}
\end{table}
\begin{table}[tbp]
\begin{center}
\topcaption{Systematic uncertainties in the $v_{4}\{\Psi_{2}\}$ values as a function of centrality in percent.
Common uncertainties are  shown at the top of the table, followed by those specific to the differential ($\pt$ dependent) and
integral ($\abs{\eta}$ dependent) measurements.}
\label{tab:sys_v4psi2_EP_combined}
\begin{scotch}{p{3cm}cccc}
\textbf{Source} &&\multicolumn{3}{c}{\textbf{Centrality}}\\
& &0--10\% &10--50\% &50--70\% \\
\hline
Particle &  & 0.5 & 0.5 & 0.5 \\
Composition \\
Centrality &  & 1.0 & 1.0 & 1.0 \\
Determination  \\
Resolution &  & 1.0 & 1.0 & 2.0 \\
Correction  \\
\hline\textbf{[Differential]}&\pt (\GeVcns)&\\ \hline
Track Quality & 0.3--0.4 &  50 &  15 &  15 \\
Requirements  & 0.4--0.8 & 6.0 & 4.0 & 4.0 \\
& 0.8--8.0 & 2.0 & 1.0 & 1.0 \\
{\textbf{Total} ${(\pt)}$ } & 0.3--0.4 &  50 &  15 &  15 \\
& 0.4--0.8 & 6.2 & 4.3 & 4.6 \\
& 0.8--8.0 & 2.5 & 1.8 & 2.5 \\
\hline \textbf{[Integral]}&${\abs{\eta}}$&\\ \hline
Track Quality & 0.0--1.6 &  10 & 5.0 & 5.0 \\
Requirements  & 1.6--2.4 &  20 & 8.0 & 8.0 \\
\hline \textbf{Total} ${(\abs{\eta})}$  & 0.0--1.6 &  10 & 5.2 & 5.5 \\
& 1.6--2.4 &  20 & 8.1 & 8.3 \\
\end{scotch}
\end{center}
\end{table}
\begin{table}[tbp]
\begin{center}
\topcaption{Systematic uncertainties in the $v_{5}\{\Psi_{5}\}$ values as a function of centrality in percent.
Common uncertainties are  shown at the top of the table, followed by those specific to the differential ($\pt$ dependent) and
integral ($\abs{\eta}$ dependent) measurements.}
\label{tab:sys_v5_EP_combined}
\begin{scotch}{p{3cm}cccc}
\textbf{Source} &&\multicolumn{3}{c}{\textbf{Centrality}}\\
& &0--10\% &10--40\% &40--50\% \\
\hline
Particle &  & 0.5 & 0.5 & 0.5 \\
Composition \\
Centrality &  & 1.0 & 1.0 & 1.0 \\
Determination  \\
Resolution &  &  10 &  10 &  10 \\
Correction  \\
\hline\textbf{[Differential]}&\pt (\GeVcns)&\\ \hline
Track Quality & 0.3--0.4 &  50 &  30 &  30 \\
Requirements  & 0.4--0.8 &  20 & 5.0 & 5.0 \\
& 0.8--8.0 & 5.0 & 3.0 & 3.0 \\
\hline \textbf{Total} ${(\pt)}$ & 0.3--0.4 &  51 &  32 &  32 \\
& 0.4--0.8 &  22 &  11 &  11 \\
& 0.8--8.0 &  11 &  11 &  11 \\
\hline \textbf{[Integral]}&${\abs{\eta}}$&\\ \hline
Track Quality &0.0--0.8&  30 &  10 &  10 \\
Requirements \\
\hline \textbf{Total} ${(\abs{\eta})}$ &0.0--0.8&  32 &  14 &  14 \\
\end{scotch}
\end{center}
\end{table}
\begin{table}[tbp]
\begin{center}
\topcaption{Systematic uncertainties in the $v_{6}\{\Psi_{6}\}$ values as a function of centrality in percent.
Common uncertainties are  shown at the top of the table, followed by those specific to the differential ($\pt$ dependent) and
integral ($\abs{\eta}$ dependent) measurements.}
\label{tab:sys_v6_EP_combined}
\begin{scotch}{p{3cm}ccc}
\textbf{Source} &&\multicolumn{2}{c}{\textbf{Centrality}}\\
& &0--10\% &10--50\% \\
\hline
Particle &  & 0.5 & 0.5 \\
Composition \\
Centrality &  & 1.0 & 1.0 \\
Determination  \\
Resolution &  &  25 &  35 \\
Correction  \\
\hline\textbf{[Differential]}&\pt (\GeVcns)&\\ \hline
Track Quality & 0.3--0.4 &  60 &  40 \\
Requirements  & 0.4--0.8 &  30 &  10 \\
& 0.8--8.0 &  10 & 5.0 \\
\hline \textbf{Total} ${(\pt)}$ & 0.3--0.4 &  65 &  53 \\
& 0.4--0.8 &  39 &  36 \\
& 0.8--8.0 &  27 &  35 \\
\hline \textbf{[Integral]}&${\abs{\eta}}$&\\ \hline
Track Quality &0.0--0.8&  40 &  15 \\
Requirements \\
\hline \textbf{Total} ${(\abs{\eta})}$  &0.0--0.8&  47 &  38 \\
\end{scotch}
\end{center}
\end{table}
\begin{table}[tbp]
\begin{center}
\topcaption{Systematic uncertainties in the $v_{6}\{\Psi_{2}\}$ values as a function of centrality in percent.
Common uncertainties are  shown at the top of the table, followed by those specific to the differential ($\pt$ dependent) and
integral ($\abs{\eta}$ dependent) measurements.}
\label{tab:sys_v6psi2_EP_combined}
\begin{scotch}{p{3cm}cccc}
\textbf{Source} &&\multicolumn{3}{c}{\textbf{Centrality}}\\
& &0--5\% &5--10\% &10--50\% \\
\hline
Particle &  & 0.5 & 0.5 & 0.5 \\
Composition \\
Centrality &  & 1.0 & 1.0 & 1.0 \\
Determination  \\
Resolution &  &  15 & 3.0 & 1.0 \\
Correction  \\
\hline\textbf{[Differential]}&\pt (\GeVcns)&\\ \hline
Track Quality & 0.3--0.4 &  60 &  40 &  40 \\
Requirements  & 0.4--0.8 &  10 &  10 & 5.0 \\
& 0.8--8.0 & 2.0 & 2.0 & 2.0 \\
\hline \textbf{Total} ${(\pt)}$  & 0.3--0.4 &  62 &  40 &  40 \\
& 0.4--0.8 &  18 &  11 & 5.2 \\
& 0.8--8.0 &  15 & 3.8 & 2.5 \\
\hline \textbf{[Integral]}&${\abs{\eta}}$&\\ \hline
Track Quality &0.0--0.8&  20 &  15 &  10 \\
Requirements \\
\hline \textbf{Total} ${(\abs{\eta})}$  &0.0--0.8&  25 &  15 &  10 \\
\end{scotch}
\end{center}
\end{table}

\begin{table}[tbp]
\begin{center}
\topcaption{Systematic uncertainties in the $v_{3}\{4\}$ values as a function of centrality in percent.
Common uncertainties are  shown at the top of the table, followed by those specific to the differential ($\pt$ dependent) and
integral ($\abs{\eta}$ dependent) measurements.}
\label{tab:sys_v3_4part_combined}
\begin{scotch}{p{3cm}ccc}
\textbf{Source} &&\multicolumn{2}{c}{\textbf{Centrality}}\\
& &10--40\% &40--60\% \\
\hline
Particle &  & 0.5 & 0.5 \\
Composition \\
Centrality &  & 1.0 & 1.0 \\
Determination  \\
Multiplicity &  & 5.0 & 5.0 \\
Fluctuations  \\
$r_{0} (\%)$&  & 5.0 & 5.0 \\
\hline\textbf{[Differential]}&\pt (\GeVcns)&\\ \hline
Track Quality & 0.3--0.5 &  20 &  10 \\
Requirements  & 0.5--0.8 &  10 & 5.0 \\
& 0.8--4.0 & 5.0 & 2.0 \\
\hline \textbf{Total} ${(\pt)}$  & 0.3--0.5 &  21 &  12 \\
& 0.5--0.8 &  12 & 8.7 \\
& 0.8--4.0 & 8.7 & 7.4 \\
\hline \textbf{[Integral]}&${\abs{\eta}}$&\\ \hline
Track Quality &0.0--0.8& 5.0 & 5.0 \\
Requirements \\
\hline \textbf{Total} ${(\abs{\eta})}$ &0.0--0.8& 8.7 & 8.7 \\
\end{scotch}
\end{center}
\end{table}
\begin{table}[tbp]
\begin{center}
\topcaption{Systematic uncertainties in the $v_{4}\{5\}$ values as a function of centrality in percent.
Common uncertainties are  shown at the top of the table, followed by those specific to the differential ($\pt$ dependent) and
integral ($\abs{\eta}$ dependent) measurements.}
\label{tab:sys_v4_5part_combined}
\begin{scotch}{p{3cm}cccc}
\textbf{Source} &&\multicolumn{3}{c}{\textbf{Centrality}}\\
& &5--10\% &10--40\% &40--60\% \\
\hline
Particle &  & 0.5 & 0.5 & 0.5 \\
Composition \\
Centrality &  & 1.0 & 1.0 & 1.0 \\
Determination  \\
Multiplicity &  & 1.0 & 2.0 & 3.0 \\
Fluctuations  \\
$r_{0} (\%)$&  & 5.0 & 3.0 & 3.0 \\
\hline\textbf{[Differential]}&\pt (\GeVcns)&\\ \hline
Track Quality & 0.3--0.5 &  15 & 5.0 & 5.0 \\
Requirements  & 0.5--0.8 &  10 & 3.0 & 3.0 \\
& 0.8--8.0 & 5.0 & 1.0 & 1.0 \\
\hline \textbf{Total} ${(\pt)}$  & 0.3--0.5 &  16 & 6.3 & 6.7 \\
& 0.5--0.8 &  11 & 4.8 & 5.3 \\
& 0.8--8.0 & 7.2 & 3.9 & 4.5 \\
\hline \textbf{[Integral]}&${\abs{\eta}}$&\\ \hline
Track Quality &0.0--0.8& 5.0 & 3.0 & 3.0 \\
Requirements \\
\hline \textbf{Total} ${(\abs{\eta})}$  &0.0--0.8& 7.2 & 4.8 & 5.3 \\
\end{scotch}
\end{center}
\end{table}

\begin{table}[tbp]
\begin{center}
\topcaption{Systematic uncertainties in the $v_{4}\{\mathrm{LYZ}\}$ values as a function of centrality in percent.
Common uncertainties are  shown at the top of the table, followed by those specific to the differential ($\pt$ dependent) and
integral ($\abs{\eta}$ dependent) measurements.}
\label{tab:sys_v4_LYZ_combined}
\begin{scotch}{p{3cm}cccc}
\textbf{Source} &&\multicolumn{3}{c}{\textbf{Centrality}}\\
& &5--10\% &10--40\% &40--60\% \\
\hline
Particle &  & 0.5 & 0.5 & 0.5 \\
Composition \\
Centrality &  & 1.0 & 1.0 & 1.0 \\
Determination  \\
Multiplicity &  & 0.1 & 0.9 & 2.0 \\
Fluctuations  \\
\hline\textbf{[Differential]}&\pt (\GeVcns)&\\ \hline
Track Quality & 0.3--0.5 &  10 & 7.0 & 3.0 \\
Requirements  & 0.5--8.0 & 3.0 & 2.0 & 1.0 \\
\hline \textbf{Total} ${(\pt)}$  & 0.3--0.5 &  10 & 7.1 & 3.8 \\
& 0.5--8.0 & 3.2 & 2.5 & 2.5 \\
\hline \textbf{[Integral]}&${\abs{\eta}}$&\\ \hline
Track Quality &0.0--0.8& 3.0 & 2.0 & 3.0 \\
Requirements \\
\hline \textbf{Total} ${(\abs{\eta})}$ &0.0--0.8& 3.2 & 2.5 & 3.8 \\
\end{scotch}
\end{center}
\end{table}
\begin{table}[tbp]
\begin{center}
\topcaption{Systematic uncertainties in the $v_{6}\{\mathrm{LYZ}\}$ values as a function of centrality in percent.
Common uncertainties are  shown at the top of the table, followed by those specific to the differential ($\pt$ dependent) and
integral ($\abs{\eta}$ dependent) measurements.}
\label{tab:sys_v6_LYZ_combined}
\begin{scotch}{p{3cm}cccc}
\textbf{Source} &&\multicolumn{3}{c}{\textbf{Centrality}}\\
& &5--10\% &10--40\% &40--60\% \\
\hline
Particle &  & 0.5 & 0.5 & 0.5 \\
Composition \\
Centrality &  & 1.0 & 1.0 & 1.0 \\
Determination  \\
Multiplicity &  & 0.1 & 0.9 & 2.0 \\
Fluctuations  \\
\hline\textbf{[Differential]}&\pt (\GeVcns)&\\ \hline
Track Quality & 0.3--0.5 &  16 &  12 & 7.5 \\
Requirements  & 0.5--8.0 & 6.0 & 4.0 & 3.0 \\
\hline \textbf{Total} ${(\pt)}$ & 0.3--0.5 &  16 &  13 & 7.8 \\
& 0.5--8.0 & 6.1 & 4.2 & 3.8 \\
\hline \textbf{[Integral]}&${\abs{\eta}}$&\\ \hline
Track Quality &0.0--0.8& 3.0 & 2.5 & 3.5 \\
Requirements \\
\hline \textbf{Total} ${(\abs{\eta})}$ &0.0--0.8& 3.2 & 2.9 & 4.2 \\
\end{scotch}
\end{center}
\end{table}

\cleardoublepage \section{The CMS Collaboration \label{app:collab}}\begin{sloppypar}\hyphenpenalty=5000\widowpenalty=500\clubpenalty=5000\textbf{Yerevan Physics Institute,  Yerevan,  Armenia}\\*[0pt]
S.~Chatrchyan, V.~Khachatryan, A.M.~Sirunyan, A.~Tumasyan
\vskip\cmsinstskip
\textbf{Institut f\"{u}r Hochenergiephysik der OeAW,  Wien,  Austria}\\*[0pt]
W.~Adam, T.~Bergauer, M.~Dragicevic, J.~Er\"{o}, C.~Fabjan\cmsAuthorMark{1}, M.~Friedl, R.~Fr\"{u}hwirth\cmsAuthorMark{1}, V.M.~Ghete, C.~Hartl, N.~H\"{o}rmann, J.~Hrubec, M.~Jeitler\cmsAuthorMark{1}, W.~Kiesenhofer, V.~Kn\"{u}nz, M.~Krammer\cmsAuthorMark{1}, I.~Kr\"{a}tschmer, D.~Liko, I.~Mikulec, D.~Rabady\cmsAuthorMark{2}, B.~Rahbaran, H.~Rohringer, R.~Sch\"{o}fbeck, J.~Strauss, A.~Taurok, W.~Treberer-Treberspurg, W.~Waltenberger, C.-E.~Wulz\cmsAuthorMark{1}
% ...
\vskip\cmsinstskip
% ...
1:~~Also at Vienna University of Technology, Vienna, Austria\\
2:~~Also at CERN, European Organization for Nuclear Research, Geneva, Switzerland\\
% ...

\end{sloppypar}
\end{document}